\documentclass[aps,prd,onecolumn,amsmath,11pt,superscriptaddress,floatfix,nofootinbib,preprintnumbers]{revtex4-2}

\usepackage{mathrsfs}
\usepackage{amsfonts}
\usepackage{amsmath,amssymb,bm}
\usepackage{array}
\usepackage{verbatim}
\usepackage{epsfig}
\usepackage{graphicx}
\usepackage{hyperref}
\hypersetup{colorlinks, linkcolor = [rgb]{0, 0, 0.5}, citecolor = [rgb]{0,0.0,0.5}, urlcolor = [rgb]{0,0.0,0.5}}
\usepackage[normalem]{ulem}
\usepackage{xcolor}
\usepackage{ulem}
\usepackage{multirow}

\usepackage{graphicx}
\graphicspath{{./figs/}}
\usepackage{dcolumn}
\usepackage{bm}
\usepackage{slashed}
\usepackage{siunitx}
\usepackage{ulem,xpatch}
\usepackage{hyperref}
\usepackage[mathlines]{lineno}
\usepackage{comment}
\usepackage{soul}
\usepackage{xcolor}
\usepackage{xfrac}

\allowdisplaybreaks[4]

\begin{document}

\title{Refined analysis of $\Omega^{-}\bar{\Omega}^{+}$ polarization in electron-positron annihilation process}

\newcommand*{\SDU}{Key Laboratory of Particle Physics and Particle Irradiation (MOE), Institute of Frontier and Interdisciplinary Science, Shandong University, Qingdao, Shandong 266237, China}\affiliation{\SDU}
\newcommand*{\HTU}{School of Physics, He'nan Normal University, Xinxiang, Henan 453007, China}\affiliation{\HTU}

\author{Zhe Zhang}\email{zhangzhe@mail.sdu.edu.cn}\affiliation{\SDU}
\author{Jiao Jiao Song}\email{songjiaojiao@ihep.ac.cn}\affiliation{\HTU}
\author{Ya-jin Zhou}\email{zhouyj@sdu.edu.cn}\affiliation{\SDU}

\begin{abstract}

We investigate the production of spin-3/2 hyperon pairs, $\Omega^-\bar{\Omega}^+$, in electron-positron annihilation within the helicity amplitude formalism. A refined selection of helicity basis matrices is proposed to relate polarization expansion coefficients and spin density matrix elements and to illuminate their inherent physical interpretations and symmetrical properties. 
With a novel parametrization scheme of helicity amplitudes, we perform an analysis of polarization correlation coefficients for double-tag $\Omega^-\bar{\Omega}^+$ pairs. We present three sets of expressions to describe the decay of $\Omega^{-}$ hyperons, and further address the existing tension in the measurements of its decay parameters, particularly $\phi_{\Omega}$. The method and the framework developed in this paper can also be applied to studies of the production and decay mechanisms of other spin-3/2 particles.

\end{abstract}

\maketitle

\newpage
\section{Introduction}
\label{s.intro}

The quantum chromodynamics (QCD) is the underlying theory of strong interactions with quarks and gluons as the fundamental degrees of freedom. Although QCD has been established for 50 years and precisely tested at high energy scales, its nonperturbative properties at low energy scales are still not well understood. The emergence of hadrons from colored quarks and gluons, or more generally the mechanism of color confinement, has become an active and challenging frontier in modern particle physics.  

The $\Omega^{-}$ hyperon, as a member of the SU(3) flavor decuplet~\cite{Gell-Mann:1962yej,Neeman:1961jhl}, plays a unique role in advancing our knowledge of the strong interaction. Its discovery~\cite{Barnes:1964pd} significantly contributed to the development of the quark model~\cite{Gell-Mann:1964ewy,Zweig:1964ruk,Zweig:1964jf} and the formulation of the color charge hypothesis~\cite{Greenberg:1964pe}. Characterized by three valence strange quarks with aligned spins and the absence of valence up or down quarks, the $\Omega^{-}$ is expected to show less relativistic effects in comparison with other octet and decuplet baryons. Yet, even after more than five decades of research, many aspects of its physical properties, apart from its charge and magnetic moment, remain largely uncharted~\cite{Beg:1964nm,Korner:1976hv,Nozawa:1990gt,Capstick:2000qj,Ramalho:2009gk,Ramalho:2019koj,Ramalho:2020laj,Jun:2021bwx,Fu:2023ijy,Zhang:2023wmd,Alexandrou:2010jv,Diehl:1991pp,Wallace:1995pf,Dobbs:2014ifa,Dobbs:2017hyd,BESIII:2020lkm,BESIII:2022kzc}. While the decuplet baryon model predicts a spin-3/2 for the $\Omega^{-}$,  direct measurements of its spin have been ongoing~\cite{Aachen-Berlin-CERN-Innsbruck-London-Vienna:1977ojz,Birmingham-CERN-Glasgow-MichiganState-Paris:1978nrw,Amsterdam-CERN-Nijmegen-Oxford:1978ylv,BaBar:2006omx}. This prediction was recently confirmed by the BESIII Collaboration~\cite{BESIII:2020lkm}. With the growing interest in spin-3/2 particles in recent years, the $\Omega^{-}$, characterized by a long lifetime and weak decay analogous to $\Lambda$ baryons, become a key focus in the study of decuplet baryons.

High-spin particles offer extensive physical insights for understanding the structure of particles and the properties of Quantum Chromodynamics (QCD) at low energies. Compared to spin-1/2 particles, which possess the spin vector and two form factors, spin-3/2 particles present a broader range of information. They include the spin vector, the rank-2 (quadrupole) spin tensor, and the rank-3 (octupole) spin tensor, totaling fifteen polarization components~\cite{PhysRev.162.1615,Doncel:1972ez,Zhao:2022lbw}, along with four form factors~\cite{Korner:1976hv}. There have been some measurements on the form factors~\cite{Dobbs:2014ifa,Dobbs:2017hyd,BESIII:2022kzc} and polarization of the Omega particle~\cite{STAR:2020xbm,BESIII:2020fqg}. However, research in these areas, especially regarding the decay parameters of $\Omega$, is not as advanced as for spin-1/2 particles, and some challenges remain.

In studies of weakly decaying particles, decay parameters like $\alpha_{D}$, $\beta_{D}$, and $\gamma_{D}$ are crucial. Specifically, the decay parameter $\alpha_{\Lambda}$ is vital for understanding the spin properties of the $\Lambda$ particle.  Recent updates in $\alpha_{\Lambda}$ measurements~\cite{BESIII:2018cnd,BESIII:2021ypr,BESIII:2022qax,BESIII:2023drj,LHCb:2020iux,Ireland:2019uja} have led to significant revisions in $\Lambda$ particle research.  Understanding the polarization of the $\Omega^{-}$ particle requires accurate measurements of $\alpha_{\Omega}$, $\beta_{\Omega}$, and $\gamma_{\Omega}$. However, these measurements are currently less precise. The most precise measurements of $\alpha_{\Omega}$  are reported by the HyperCP Collaboration~\cite{HyperCP:2005wwr,HyperCP:2005ukv}. Prior to 2021, there were no direct measurements of $\beta_{\Omega}$ and $\gamma_{\Omega}$. It was commonly assumed that $\gamma_{\Omega}$ would be either $+1$ or $-1$~\cite{Luk:1988as,Kim:1992az}. However, recent measurements have presented conflicting results. The STAR Collaboration identified $\gamma_{\Omega}$ as $1$~\cite{STAR:2020xbm}, while the BESIII Collaboration reported it to be approximately $-0.5$~\cite{BESIII:2020lkm}, challenging previous assumptions and findings. It is important to acknowledge the significant measurement uncertainties in previous experiments. There is a need for more precise measurements to determine the decay parameters of the $\Omega$ particle.

Furthermore, the differences in decay parameters between particles and antiparticles are directly linked to the asymmetry between matter and antimatter in the universe. Extensive research on CP violation in the decays of spin-1/2 particles, including $\Lambda$~\cite{BESIII:2018cnd,BESIII:2021ypr,BESIII:2022qax,BESIII:2023drj}, $\Xi$~\cite{BESIII:2021ypr,BESIII:2022lsz,BESIII:2023drj}, and $\Sigma$~\cite{STAR:2020xbm}, has been conducted by the BESIII Collaboration. However, no evidence of CP violation beyond the Standard Model has been found. Investigating CP violation in the decays of higher spin and strangeness particles, such as $\Omega^{-}$, is an essential and ongoing area of research.

The  $e^{+}e^{-} \rightarrow \Omega^{-}\bar{\Omega}^{+}$ proccess is a key to investigating the form factors and decay parameters of the $\Omega^{-}$ particle. Using the $\psi(3686)$ dataset of $\left(448.1 \pm 2.9\right) \times 10^{6}$ events~\cite{Ablikim:2012pj,BESIII:2017tvm}, along with theoretical helicity formulations about this process~\cite{Perotti:2018wxm}, the BESIII collaboration successfully  achieved the first measurement of the three decay parameters of the $\Omega^{-}$ particle~\cite{BESIII:2020lkm}. However, these measurements faced considerable uncertainties, largely attributed to the limited size of the dataset and significant background noise in single-tag measurements. The recent accumulation of a larger $\psi(3686)$ dataset by the BESIII Collaboration, estimated about $\left(27.08 \pm 0.14\right) \times 10^8$ events~\cite{BESIII:2023olq}, opens up opportunities for more precise measurements of the $\Omega^{-}$ decay parameters. The adoption of double-tag $\Omega^{-}\bar{\Omega}^{+}$ decay chains in the measurement is expected to reduce background noise, representing a methodological advancement.
 
In this paper, we begin with an analysis of the $\Omega^{-}$ particles within the helicity formalism, building on previous studies~\cite{Jacob:1959at,Korner:1976hv,Chen:2007zzf,Perotti:2018wxm}. Within this framework, we expand the complete information of polarization properties and form factors for the $\Omega^{-}$ particles into four complex amplitudes: $H_{1}$, $H_{2}$, $H_{3}$, and $H_{4}$. We review methods for decomposing polarization components in the helicity formalism and the general spin density matrix formalism.  To bridge the spin components in helicity formalism with those in the spin density matrix, we introduce a new set of basis matrices in the helicity framework. This approach enables a clear understanding of the physical interpretation of the polarization components of the $\Omega^{-}$ particle. 
 
We propose a parametrization scheme for the helicity amplitudes of spin-3/2 particles, drawing an analogy with the helicity amplitude parametrization of spin-1/2 particles. This scheme includes six parameters: $\alpha_\psi$, $\alpha_1$, $\alpha_2$, $\phi_1$, $\phi_3$, and $\phi_4$. This parametrization simplifies the relationships between these parameters and various polarization coefficients. For instance, the angular dependence of cross-section terms is solely related to $\alpha_\psi$. We also identify the ranges for these parameters. Then, we present the complete set of polarization coefficients for $\Omega$ particles: 7 nonzero coefficients for single-tag $\Omega^{-}$ and 116 nonzero coefficients for double-tag $\Omega^{-}\bar{\Omega}^{+}$. Using  the physical interpretation of polarization components, we investigate the survival of these coefficients and analyze their behavior under parity and CP transformations. Notably, we identify specific parameters solution sets that zero polarization in single-tag $\Omega^{-}$ case and minimize polarization correlations in double-tag $\Omega^{-}\bar{\Omega}^{+}$ system. Furthermore, through an analysis of the parameter value ranges and their corresponding expressions for polarization correlation coefficients, we establish the boundaries of these coefficients.  These analyses provide fresh insights into the polarization dynamics involved in the production of $\Omega^{-}\bar{\Omega}^{+}$ in positron-electron annihilation processes.

We introduce three formalisms to describe the decay of the $\Omega^{-}$ particle, characterized by three decay parameters: $\alpha_{\Omega}$, $\beta_{\Omega}$, and $\gamma_{\Omega}$. While these formalisms are equivalent, they offer unique insights on the decay of spin-3/2 particles.  We address the challenges associated with measurements of the decay parameter $\phi_{\Omega}$ for the $\Omega^{-}$ particle. Using the maximum likelihood method, we assess the sensitivity of $\phi_{\Omega}$ to the number of observed events, $N$, in both single-tag and double-tag cases. Our findings reveal that double-tag measurements provide statistical advantages and effectively reduce background noise. With the accumulated data from the BESIII Collaboration on $\psi(3686)$ events, the statistical sensitivity for $\phi_{\Omega}$ in double-tag measurements is expected to reach approximately 2\%. This level of precision has the potential to resolve the current discrepancies in the measurements of decay parameters for the $\Omega$ particle.

Our analysis focuses on the $e^{+}e^{-}\rightarrow\Omega^{-}\bar{\Omega}^{+}$ process, while the framework we developed could extend to other spin-3/2 particle production processes like $e^{+}e^{-}\rightarrow\Xi^{-}(1530)\bar{\Xi}^{+}(1530)$, $e^{+}e^{-}\rightarrow\Sigma^{-}(1385)\bar{\Sigma}^{+}(1385)$, and so forth.

The organization of this paper is as follows. In Sec.~\ref{s.production}, we present the production density matrix for $\Omega^{-}\bar{\Omega}^{+}$ in $e^+e^-$ annihilation within the helicity formalism. In Sec.~\ref{s.sdm}, we  introduce a novel set of helicity basis matrices and establish the relationships between various conventions of spin components decomposition. In Sec.~\ref{s.spin_analysis}, we propose a parametrization scheme for helicity amplitudes, and further present detailed analyses of single-tag $\Omega^{-}$ polarization expansion coefficients and double-tag $\Omega^{-}\bar{\Omega}^{+}$ polarization correlation coefficients.  In Sec.~\ref{s.decay}, we presents three equivalent formulations for $\Omega^{-}$ decay. In Sec~\ref{s.sensitivity}, we forecast the statistical sensitivity for the decay parameter $\phi_{\Omega}$  at BESIII via the maximum likelihood approach. Finally, in Sec.~\ref{s.summary}, we provide a brief summary.

\section{$\Omega^{-}\bar{\Omega}^{+}$ production in $e^{+}e^{-}$ annihilation}\label{s.production}

\begin{figure}
\begin{centering}
\includegraphics[width=0.8\textwidth]{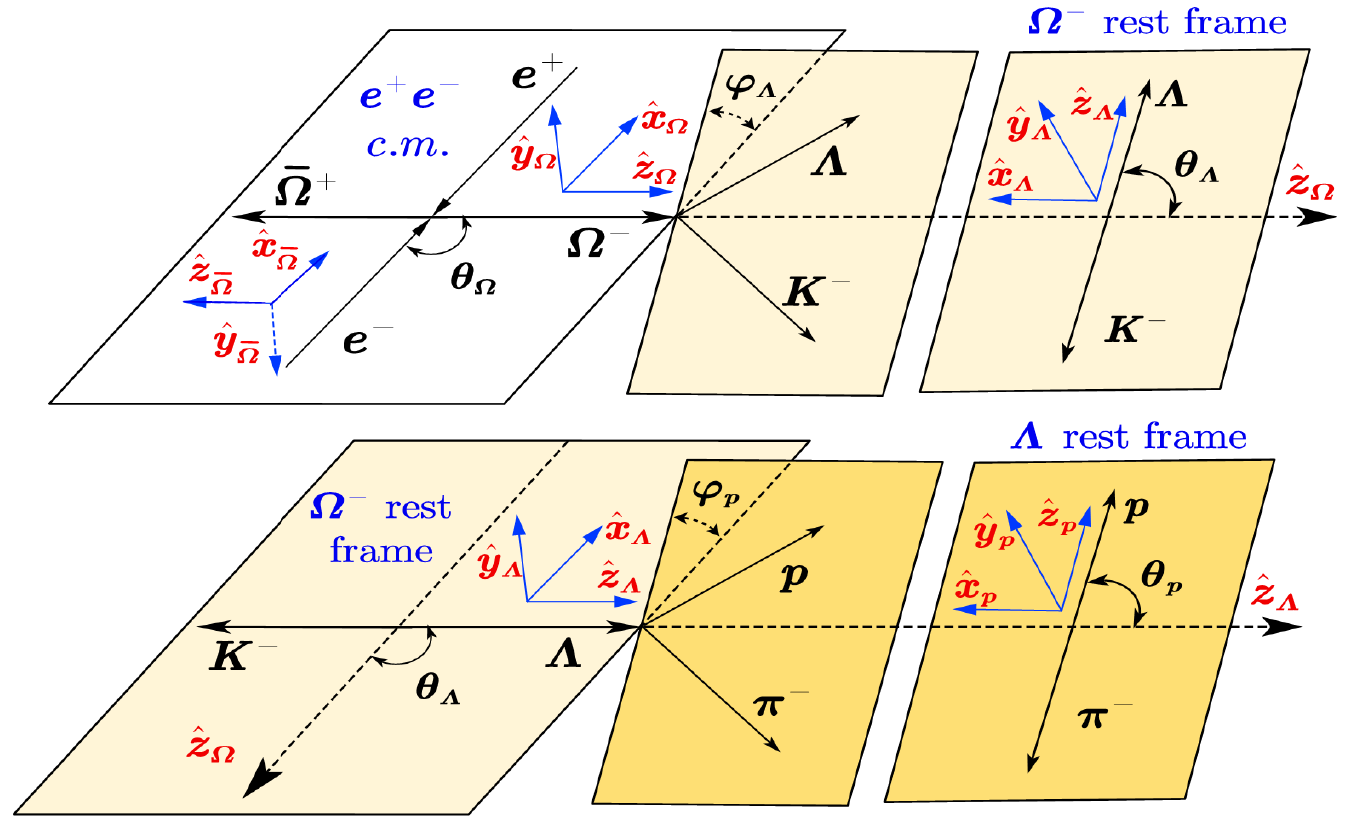}
\par\end{centering}
\caption{\label{fig:Omega}Definition of helicity formalism coordinate systems and angles. The angles  $\theta_{\Lambda}$, $\phi_{\Lambda}$, $\theta_{p}$, and $\phi_{p}$
represent the polar and azimuthal angles of the $\Lambda$ and the proton, in the $\Omega^{-}$ rest frame and the $\Lambda$ rest frame, respectively.}

\end{figure}

The $e^{+}e^{-}\rightarrow\Omega^{-}\bar{\Omega}^{+}$ process serves as a crucial avenue for probing $\Omega^{-}$ particle characteristics, with experiments conducted at multiple facilities such as $BABAR$~\cite{BaBar:2001yhh}, BESIII~\cite{BESIII:2009fln,BESIII:2020nme}, Belle II~\cite{Belle-II:2018jsg}, and the proposed STCF~\cite{Achasov:2023gey}. In this section, we present the production density matrix for $\Omega^{-}\bar{\Omega}^{+}$ in $e^{+}e^{-}$ annihilation process within the helicity formalism. We detail the coordinate system and define the angles relevant to the $\Omega^{-}\bar{\Omega}^{+}$ production, as illustrated in Fig.~\ref{fig:Omega}. Given that particle polarization is typically examined via decay processes, we establish the coordinate system and angles necessary to analyze the cascade decay of the  $\Omega^{-}$ particle, adhering to the same formalism.

In the helicity formalism, the production of $\Omega^{-}\bar{\Omega}^{+}$ in $e^{+}e^{-}$ annihilation is analyzed in the center-of-mass ($c.m.$) frame of $e^{+}e^{-}$. The polar angle $\theta_{\Omega}$ is defined by the angle between the $\Omega^{-}$ particle and the positron. In the $\Omega^{-}$ coordinate system, where we decompose the polarization components of $\Omega^{-}$, the $\hat{z}_{\Omega}$ axis aligns with the momentum of the $\Omega^{-}$ particle. The $\hat{y}_{\Omega}$ axis, orthogonal to the momenta of $e^{+}$ and $\Omega^{-}$, is defined by the cross product $\hat{y}_{\Omega} = \vec{p}_{e^{+}} \times \vec{p}_{\Omega}$. The $\hat{x}_{\Omega}$ axis is determined using the right-hand rule. The coordinate system for $\bar{\Omega}^{+}$ mirrors this, but uses the momentum of $\bar{\Omega}^{+}$ instead.
The back-to-back production of $\Omega^{-}$ and $\bar{\Omega}^{+}$, where $\vec{p}_{\bar{\Omega}} = -\vec{p}_{\Omega}$, implies the following relations
\begin{align}
\hat{z}_{\bar{\Omega}}=-\hat{z}_{\Omega},\quad\hat{y}_{\bar{\Omega}}=-\hat{y}_{\Omega},\quad\hat{x}_{\bar{\Omega}}=\hat{x}_{\Omega}.\label{eq:Omega_Omegabar_axis}
\end{align}

We describe the decay of the $\Omega^{-}$ particle in its rest frame, where the $x_{\Omega}$-$y_{\Omega}$-$z_{\Omega}$ coordinate system is remained.  In this frame, the decay angles of the $\Lambda$ particle are described by its polar and azimuthal angles. The coordinate system for the $\Lambda$ particle is then established as follows: the $\hat{z}_{\Lambda}$ aligns with the  momentum of the $\Lambda$ particle; the $\hat{y}_{\Lambda}$, perpendicular to the direction of $z_{\Omega}$ and the momentum of  the $\Lambda$, is defined as $\hat{y}_{\Lambda} = \vec{z}_{\Omega} \times \vec{p}_{\Lambda}$; the $\hat{x}_{\Lambda}$ is determined using the right-hand rule.  The coordinate system of proton, set in the $\Lambda$ rest frame, adheres to similar principles.  Likewise, the decay coordinate systems for the $\bar{\Omega}^{+}$ side mirrors those of $\Omega^{-}$ side. In this mirrored system, the momenta of the $\Omega^{-}$, $\Lambda$, and proton are replaced by the corresponding momenta of the $\bar{\Omega}^{+}$, $\bar{\Lambda}$, and antiproton, respectively.

With the defined coordinate systems and angles, the production density matrix for the $e^{+}e^{-}\to\Omega^{-}\bar{\Omega}^{+}$ process is presented as~\cite{Perotti:2018wxm},
\begin{align}\label{eq:production density matrix}
\rho_{B_{1}\bar{B}_{2}}^{\lambda_{1},\lambda_{2};\lambda_{1}^{\prime},\lambda_{2}^{\prime}} \propto&  A_{\lambda_{1},\lambda_{2}}A_{\lambda_{1}^{\prime},\lambda_{2}^{\prime}}^{*}\rho_{1}^{\lambda_{1}-\lambda_{2},\lambda_{1}^{\prime}-\lambda_{2}^{\prime}},
\end{align}
where $A_{\lambda_{1},\lambda_{2}}$ and $A_{\lambda_{1}^{\prime},\lambda_{2}^{\prime}}^{}$ are the transition amplitudes with the helicities $\lambda_{1}$, $\lambda_{1}^{\prime}$ for $\Omega^{-}$ and $\lambda_{2}$, $\lambda_{2}^{\prime}$ for $\bar{\Omega}^{+}$. The matrix $\rho_{1}$ is defined as
\begin{align}
\rho_{1}^{i,j}\left(\theta_{\Omega}\right)=  \sum_{\kappa=\pm1}D_{\kappa,i}^{1*}\left(0,\theta_{\Omega},0\right) D_{\kappa,j}^{1*}\left(0,\theta_{\Omega},0\right),
\end{align}
 where $D_{\kappa,i}^J(0,\theta_{\Omega},0)$ represents the Wigner D-matrix, $\theta_{\Omega}$ is the helicity angle of the $\Omega^{-}$ particle, $\kappa$ denotes the helicity difference between the initial $e^{+}$ and $e^{-}$ states, constrained to $\pm1$ when the electron mass is negligible. For unpolarized lepton beams, a straightforward summation over $\kappa$ is suitable.

Considering the principles of parity conservation and charge conjugation invariance, only four independent transition amplitudes remain relevant~\cite{Perotti:2018wxm},
\begin{align}\label{eq:helicity amplitudes}
\begin{aligned}
H_{1}= & A_{1/2,1/2}=A_{-1/2,-1/2},\\
H_{2}= & A_{1/2,-1/2}=A_{-1/2,1/2},\\
H_{3}= & A_{3/2,1/2}=A_{-3/2,-1/2}\\
= & A_{1/2,3/2}=A_{-1/2,-3/2},\\
H_{4}= & A_{3/2,3/2}=A_{-3/2,-3/2}.
\end{aligned}
\end{align}
The transition amplitudes matrix is given by
\begin{align}
A_{i,j}=\left(\begin{array}{cccc}
H_{4} & H_{3} & 0 & 0\\
H_{3} & H_{1} & H_{2} & 0\\
0 & H_{2} & H_{1} & H_{3}\\
0 & 0 & H_{3} & H_{4}
\end{array}\right).\label{eq:transition matrix}
\end{align}
This matrix encompasses information of the form factors and polarization characteristics of the $\Omega^{-}$ baryon. Detailed connections between helicity amplitudes and form factors are presented in Appendix~\ref{A.form_factors}, guiding in both lattice QCD and quark models researches on its structural properties.

In our analysis of the $e^{+}e^{-} \rightarrow \Omega^{-}\bar{\Omega}^{+}$ process, we focus on the electromagnetic and strong interactions where parity conservation applies.  Even though parity and CP violation effects are theoretically predicted to be detectable in the $e^{+}e^{-} \rightarrow \Lambda\bar{\Lambda}$ process \cite{He:2022jjc}, we do not include these aspects in our analysis due to the markedly lower production rates of $\Omega^{-}$ compared to $\Lambda$.

\section{Spin density matrix}\label{s.sdm}

In this paper, we analyze the polarization states of particles using spin density matrices in their rest frames, focusing on particles with spins of 1/2 and 3/2. Various methods are available for decomposing polarization components, including the helicity formalism~\cite{Perotti:2018wxm} and the general spin density matrix formalism~\cite{PhysRev.162.1615,Zhao:2022lbw,Zhang:2023wmd}. Although these methods are fundamentally similar, their varying conventions often cause confusion. To address the confusion, we  bridge the spin components in helicity formalism with those in the general spin density matrix. 

We now review the polarization components decomposition in various formalisms, starting with the general spin density matrix formalism due to its clear physical interpretations of spin components. For a spin-1/2 particle, the polarization states are represented by a $2 \times 2$ Hermitian matrix in a Cartesian coordinate system, formulated as
\begin{align}
\rho_{1/2}= & \frac{1}{2}\left(\bm{1}+S^{i}\sigma^{i}\right),\label{eq:spin-1/2}
\end{align}
where $\sigma^{i}$ denotes the Pauli matrices, and $S^{i}$, the rank-one spin vector, comprises
\begin{align}
S^{i}= & \left(S_{x},S_{y},S_{z}\right).\label{eq:vector_spin_1/2}
\end{align}
This spin density matrix formulation adheres to the normalization condition $\text{Tr}[\rho] = 1$.

For spin-3/2 particles, the spin density matrix is represented as~\cite{Zhao:2022lbw,Zhang:2023wmd}
\begin{align}
\rho= & \frac{1}{4}\left(\bm{1}+\frac{4}{5}S^{i}\Sigma^{i}+\frac{2}{3}T^{ij}\Sigma^{ij}+\frac{8}{9}R^{ijk}\Sigma^{ijk}\right),\label{eq:spin-3/2}
\end{align}
where  $\Sigma^{i}$, $\Sigma^{ij}$, and $\Sigma^{ijk}$ are independent, orthonormal, and Hermitian basis matrices. The matrices $\Sigma^{i}$, in the $S_{z}$ representation, are defined as
\begin{align}
\Sigma^{x}=\frac{1}{2}\left(\begin{array}{cccc}
0 & \sqrt{3} & 0 & 0\\
\sqrt{3} & 0 & 2 & 0\\
0 & 2 & 0 & \sqrt{3}\\
0 & 0 & \sqrt{3} & 0
\end{array}\right),\quad\Sigma^{y}=\frac{i}{2}\left(\begin{array}{cccc}
0 & -\sqrt{3} & 0 & 0\\
\sqrt{3} & 0 & -2 & 0\\
0 & 2 & 0 & -\sqrt{3}\\
0 & 0 & \sqrt{3} & 0
\end{array}\right),\quad\Sigma^{z}=\frac{1}{2}\left(\begin{array}{cccc}
3 & 0 & 0 & 0\\
0 & 1 & 0 & 0\\
0 & 0 & -1 & 0\\
0 & 0 & 0 & -3
\end{array}\right).\label{eq:spin_vector}
\end{align}
The rest of of the basis matrices can be formulated using the $\Sigma^{i}$ matrices,
\begin{align}
\Sigma^{ij}= & \frac{1}{2}\left(\Sigma^{i}\Sigma^{j}+\Sigma^{j}\Sigma^{i}\right)-\frac{5}{4}\delta^{ij}\bm{1},\label{eq:spin_tensor1}\\
\Sigma^{ijk}= & \frac{1}{6}\Sigma^{\{i}\Sigma^{j}\Sigma^{k\}}-\frac{41}{60}\left(\delta^{ij}\Sigma^{k}+\delta^{jk}\Sigma^{i}+\delta^{ki}\Sigma^{j}\right).\label{eq:spin_tensor2}
\end{align}
The polarization characteristics of particles are described by the components $S^{i}$, $T^{ij}$, and $R^{ijk}$, representing the spin vector, the rank-2 spin tensor, and the rank-3 spin tensor, respectively, which include
\begin{align}
S^{i} & :S_{L},S_{T}^{x},S_{T}^{y},\label{eq:spin_componences_1}\\
T^{ij} & :S_{LL},S_{LT}^{x},S_{LT}^{y},S_{TT}^{xx},S_{TT}^{xy},\\
R^{ijk} & :S_{LLL},S_{LLT}^{x},S_{LLT}^{y},S_{LTT}^{xx},S_{LTT}^{xy},S_{TTT}^{xxx},S_{TTT}^{yxx}.
\label{eq:spin_componences_2}
\end{align}
These 15 independent polarization components reveal specific eigenstate probabilities in the spin density matrix $\rho_{3/2}$.   Detailed probabilistic interpretations of these components can be found in Appendix~\ref{B.Probabilistic}. While our definitions generally align with those in Refs.~\cite{Zhao:2022lbw,Zhang:2023wmd}, we introduce a notable variation in the definition of $S_{LTT}^{xy}$ for a refined analysis of polarization symmetries. The domains of these components are not normalized to $1$, as outlined in Appendix ~\ref{B.Probabilistic}. 

In the helicity formalism, spin density matrices are concise yet may lack clear probabilistic interpretations.  To address this, linking these polarization coefficients with the corresponding elements in the general spin density matrix representation offers a direct understanding.

The spin density matrix for spin-1/2 particles in the helicity formalism is typically represented as~\cite{Perotti:2018wxm}
\begin{align}
\rho_{1/2}^{B} & =\frac{1}{2}\sum_{\mu}I_{\mu}\sigma_{\mu},\label{eq:spin-1/2_helicity}
\end{align}
where $\mu$ runs through $0$, $x$, $y$, and $z$. The term $\sigma_{0}$ represents the $2\times2$ identity matrix, and $\sigma_{x}$, $\sigma_{y}$, and $\sigma_{z}$ are the Pauli matrices. The trace $\text{Tr}[\rho]$ equals $I_{0}$, representing the total cross-section, which typically is not normalized to $1$. Consequently, the spin vector of the particle is defined as $\vec{S}=\vec{I}/I_{0}$.

The spin density matrix for spin-3/2 particles is typically represented as~\cite{Perotti:2018wxm,Zhang:2023wmd}
\begin{align}
\rho_{3/2}^{B} & =\sum_{\mu=0}^{15}S_{\mu}\Sigma_{\mu},\label{eq:spin-3/2_helicity}
\end{align}
where $\Sigma_{0},\Sigma_{1},...,\Sigma_{15}$ form a complete set of orthogonal basis matrices.   The coefficients $S_{0}, S_{1}, ..., S_{15}$ correspond to the polarization components of the particle.  We adopt the basis matrices framework in Ref.~\cite{Zhang:2023wmd}, as explained in Appendix~\ref{C.Matrices}. In this framework, we establish a direct correlation between $S_{0}, S_{1}, ..., S_{15}$ and the spin components defined in Eqs.~\eqref{eq:spin_componences_1}-\eqref{eq:spin_componences_2}.  This relationship is detailed in Table~\ref{tab:correspondence}.

\begin{table}[ht]
\caption{\label{tab:correspondence}The correspondence between the expansion
coefficients $S_0, S_1, ..., S_{15}$ and the spin components in Eqs.~\eqref{eq:spin_componences_1}-\eqref{eq:spin_componences_2}.  Dividing the first-row coefficients by $S_{0}$ provides the second-row spin components, exemplified by $S_{1}/S_{0}=S_{L}$.}
\renewcommand\arraystretch{1.5}
\begin{tabular}{c|c|c|c|c|c|c|c|c|c|c|c|c|c|c|c}
\hline\hline  ~~\ensuremath{S_{0}}~~  &  ~~\ensuremath{S_{1}}~~  &  ~~\ensuremath{S_{2}}~~  &  ~~\ensuremath{S_{3}}~~  &  ~~\ensuremath{S_{4}}~~  &  ~~\ensuremath{S_{5}}~~  &  ~~\ensuremath{S_{6}}~~  &  ~~\ensuremath{S_{7}}~~  &  ~~\ensuremath{S_{8}}~~  &  ~~\ensuremath{S_{9}}~~  &  ~~\ensuremath{S_{10}}~~  &  ~~\ensuremath{S_{11}}~~  &  ~~\ensuremath{S_{12}}~~  &  ~~\ensuremath{S_{13}}~~  &  ~~\ensuremath{S_{14}}~~  &  ~~\ensuremath{S_{15}}~~\\
\hline  \ensuremath{1}  &  \ensuremath{S_{L}}  &  \ensuremath{S_{T}^{x}}  &  \ensuremath{S_{T}^{y}}  &  \ensuremath{S_{LL}}  &  \ensuremath{S_{LT}^{x}}  &  \ensuremath{S_{LT}^{y}}  &  \ensuremath{S_{TT}^{xx}}   & \ensuremath{S_{TT}^{xy}}  &  \ensuremath{S_{LLL}}  &  \ensuremath{S_{LLT}^{x}}  &  \ensuremath{S_{LLT}^{y}}  &  \ensuremath{S_{LTT}^{xx}}  &  \ensuremath{S_{LTT}^{xy}}  &  \ensuremath{S_{TTT}^{xxx}}  &  \ensuremath{S_{TTT}^{yxx}}
\\\hline\hline \end{tabular}
\end{table}

\section{Spin analysis of $\Omega$ baryons} \label{s.spin_analysis}

In this section, we propose a new parametrization scheme for helicity amplitudes to enhance our understanding of $\Omega^{-}$ polarization. Then we analyze the polarization of single-tag $\Omega^{-}$ and provide comprehensive expressions for the polarization correlation coefficients of double-tag $\Omega^{-}\bar{\Omega}^{+}$. By analyzing the polarization correlation coefficients constrained by  the ranges of the helicity parameters, we obtain the boundaries for these coefficients. We identify specific parameters solution sets that result in zero polarization in single-tag $\Omega^-$ case and minimize polarization correlations in double-tag $\Omega^{-}\bar{\Omega}^{+}$ system. We also delve into the inherent symmetries of these coefficients  by examining their physical interpretations.

\subsection{Parametrization scheme} \label{s.parametrization}
The polarization of the $\Omega^{-}$ particle is characterized by four helicity amplitudes, detailed in Eq.~\eqref{eq:helicity amplitudes}. In Ref.~\cite{BESIII:2020lkm} a simple parametrization is used to relate these amplitudes with helicity parameters $h_i$ and $\phi'_i$ by defining $H_1/H_2=h_1 e^{i\phi'_1}$, $H_3/H_2=h_3 e^{i\phi'_3}$, $H_4/H_2=h_4 e^{i\phi'_4}$, and obtained two sets of fit values of the helicity parameters, called Solution I and Solution II. This parametrization is straightforward, but the polarization information is mixed in these parameters. We aim to introduce a new parametrization scheme to clear the polarization interpretation of helicity parameters. To align with the parametrization for spin-1/2 particles as in Refs.~\cite{Chen:2007zzf, Perotti:2018wxm, Faldt:2017kgy}, we start with the introduction of a parameter $\alpha_{\psi}$, and represent the production cross-section of $\Omega^-\bar{\Omega}^+$ as $d\sigma \propto 1 + \alpha_{\psi}\cos^2\theta_{\Omega}$. Our approach also simplifies the connections between different polarization components and parameters.  By examining the expressions for single-tag $\Omega^{-}$ polarization in Appendix~\ref{D.spin_correlations}, we establish our parametrization  scheme as following
\begin{align}
H_{1} =& \frac{1}{2\sqrt{2}}\sqrt{1-\alpha_{\psi}-\alpha_{1}}\exp\left[i\left(\phi_{1}+\phi_{3}\right)\right],\nonumber\\
H_{2} =& \frac{1}{\sqrt{2}}\sqrt{\alpha_{2}},\nonumber\\
H_{3} =& \frac{1}{2}\sqrt{1+\alpha_{\psi}-\alpha_{2}}\exp\left[i\phi_{3}\right],\nonumber\\
H_{4} =& \frac{1}{2\sqrt{2}}\sqrt{1-\alpha_{\psi}+\alpha_{1}}\exp\left[i\left(\phi_{4}+\phi_{3}\right)\right],
\label{eq:parametrization scheme}
\end{align}
where the domains of these helicity parameters are,
\begin{align}
-1\leq & \alpha_{\psi}  \leq1,\label{eq:range1}\\
-1+\alpha_{\psi}\leq & \alpha_{1}  \leq1-\alpha_{\psi},\\
0\leq & \alpha_{2}  \leq1+\alpha_{\psi}.\label{eq:range2}
\end{align} 
These domains simplify the experimental constraints on the parameters, and facilitate the analysis of $\Omega^{-}$ spin properties. 
\begin{table}[ht]
\caption{\label{tab:parametrization}The new parametrization scheme calculated using Eq.~\eqref{eq:parameters_comparition}. The data sets are from the measurements of the $e^{+}e^{-}\rightarrow\gamma^{*}\rightarrow\psi\left(3686\right)\rightarrow\Omega^{-}\bar{\Omega}^{+}$ process by the BESIII collaboration~\cite{BESIII:2020lkm}.}
\renewcommand\arraystretch{1.5}
\[
\begin{tabular}{c|c|c}
\hline\hline ~~ Parameter~~  &  ~~~~~Solution I~~~~~  &  ~~~~~Solution II~~~~~ \\
\hline  \ensuremath{\alpha_{\psi}}  &  0.237\ensuremath{\pm}0.109  &  0.233\ensuremath{\pm}0.095\\
 \ensuremath{\alpha_{1}}  &  -0.371\ensuremath{\pm}0.202  &  -0.353\ensuremath{\pm}0.175\\
 \ensuremath{\alpha_{2}}  &  1.090\ensuremath{\pm}0.128  &  1.076\ensuremath{\pm}0.116\\
 \ensuremath{\phi_{1}}  &  4.37\ensuremath{\pm}0.44  &  6.09\ensuremath{\pm}0.44\\
 \ensuremath{\phi_{3}}  &  2.60\ensuremath{\pm}0.18  &  2.57\ensuremath{\pm}0.17\\
 \ensuremath{\phi_{4}}  &  4.02\ensuremath{\pm}0.89  &  5.08\ensuremath{\pm}0.70
\\\hline\hline \end{tabular}
\]
\end{table}

We provide detailed relations between these helicity parameters and those in Ref.~\cite{BESIII:2020lkm}, 
\begin{align}
\alpha_{\psi} & =\frac{1-2h_{1}^{2}+2h_{3}^{2}-2h_{4}^{2}}{1+2h_{1}^{2}+2h_{3}^{2}+2h_{4}^{2}},\nonumber\\
\alpha_{1} & =-\frac{4\left(h_{1}^{2}-h_{4}^{2}\right)}{1+2h_{1}^{2}+2h_{3}^{2}+2h_{4}^{2}},\nonumber\\
\alpha_{2} & =\frac{2}{1+2h_{1}^{2}+2h_{3}^{2}+2h_{4}^{2}},\nonumber\\
\phi_{1} & =\phi_{1}^{\prime}-\phi_{3}^{\prime}+2\pi N,\nonumber\\
\phi_{3} & =\phi_{3}^{\prime}+2\pi N,\nonumber\\
\phi_{4} & =\phi_{4}^{\prime}-\phi_{3}^{\prime}+2\pi N,
\label{eq:parameters_comparition}
\end{align}
where $N$ is an arbitrary integer. Using these relations we obtain the helicity parameters in our scheme corresponding to the two sets of fit values in Ref.~\cite{BESIII:2020lkm}, and list them in Table~\ref{tab:parametrization} for later use.

\subsection{Single-tag $\Omega^{-}$} \label{s.single}

To analyze the polarization of the single-tag $\Omega^{-}$, we sum over the spin states of its counterpart, $\bar{\Omega}^{+}$.   Using Eq.~\eqref{eq:production density matrix}, the production density matrix for a single-tag $\Omega^{-}$ is described as~\cite{Zhang:2023wmd}
\begin{align}
\rho_{\Omega}\propto\sum_{\lambda_{2}}A_{\lambda_{1},\lambda_{2}}A_{\lambda_{1}^{'},\lambda_{2}}^{*}\rho_{1}^{\lambda_{1}-\lambda_{2},\lambda_{1}^{'}-\lambda_{2}}(\theta_{\Omega}).
\label{eq:production_density_matrix_Omega}
\end{align}
Inserting the transition amplitude $A_{\lambda_i,\lambda_j}$, which expressed in terms of the helicitiy parameters defined in the previous subsection, results
\begin{align}
\rho_{\Omega} = & \left(\begin{array}{cccc}
m_{11} & c_{12} & c_{13} & 0\\
c_{12}^{*} & m_{22} & im_{23} & c_{13}^{*}\\
c_{13}^{*} & -im_{23} & m_{22} & -c_{12}^{*}\\
0 & c_{13} & -c_{12} & m_{11}
\end{array}\right),\label{eq:spin_density_matrix_Omega}
\end{align}
where $m_{11}$, $m_{22}$, $m_{23}$ are real, and $c_{12}$, $c_{13}$ are complex functions, given by,
\begin{align}
m_{11} & =  \frac{1}{8}\left[\left(2+\alpha_{1}-\alpha_{2}\right)+\left(2\alpha_{\psi}-\alpha_{1}-\alpha_{2}\right)\cos^{2}\theta_{\Omega}\right],\\
m_{22} & =  \frac{1}{8}\left[\left(2-\alpha_{1}+\alpha_{2}\right)+\left(2\alpha_{\psi}+\alpha_{1}+\alpha_{2}\right)\cos^{2}\theta_{\Omega}\right],\\
m_{23} & =  \frac{\sqrt{2}}{8}\sqrt{\alpha_{2}\left(1-\alpha_{\psi}-\alpha_{1}\right)}\sin2\theta_{\Omega}\sin\left(\phi_{1}+\phi_{3}\right),\\
c_{12} & =  -\frac{1}{16}\sqrt{1+\alpha_{\psi}-\alpha_{2}}\sin2\theta_{\Omega}\nonumber\\
&\quad\times\left[\sqrt{1-\alpha_{\psi}-\alpha_{1}}\left(\cos\phi_{1}-i\sin\phi_{1}\right)-\sqrt{1-\alpha_{\psi}+\alpha_{1}}\left(\cos\phi_{4}+i\sin\phi_{4}\right)\right],\\
c_{13} & =  \frac{\sqrt{2}}{8}\sqrt{\alpha_{2}\left(1+\alpha_{\psi}-\alpha_{2}\right)}\sin^{2}\theta_{\Omega}\left(\cos\phi_{3}+i\sin\phi_{3}\right).
\end{align}
By comparing Eq.~\eqref{eq:spin_density_matrix_Omega} and Eq.~\eqref{eq:spin-3/2_helicity} we can express the polarization components $S_i$ in terms of the helicity parameters, and list the non-zero terms as follows,
\begin{align}
S_{0} = & 1+\alpha_{\psi}\cos^{2}\theta_{\Omega},\label{eq:S0}\\
S_{3} = & -\frac{1}{8}\sin2\theta_{\Omega}\left[\sqrt{3}\sqrt{1+\alpha_{\psi}-\alpha_{2}}\left(\sqrt{1-\alpha_{\psi}-\alpha_{1}}\sin\phi_{1}+\sqrt{1-\alpha_{\psi}+\alpha_{1}}\sin\phi_{4}\right)\right.\nonumber\\
       & \left.+2\sqrt{2}\sqrt{\alpha_{2}\left(1-\alpha_{\psi}-\alpha_{1}\right)}\sin\left(\phi_{1}+\phi_{3}\right)\right],\\
S_{4} = & \frac{1}{2}\left[\left(\alpha_{1}-\alpha_{2}\right)-\left(\alpha_{1}+\alpha_{2}\right)\cos^{2}\theta_{\Omega}\right],\\
S_{5} = & -\frac{\sqrt{3}}{4}\sin2\theta_{\Omega}\sqrt{1+\alpha_{\psi}-\alpha_{2}}\left(\sqrt{1-\alpha_{\psi}-\alpha_{1}}\cos\phi_{1}-\sqrt{1-\alpha_{\psi}+\alpha_{1}}\cos\phi_{4}\right),\\
S_{7} = & \frac{\sqrt{6}}{2}\sin^{2}\theta_{\Omega}\sqrt{\alpha_{2}\left(1+\alpha_{\psi}-\alpha_{2}\right)}\cos\phi_{3},\\
S_{11}= &-\frac{1}{20}\sin2\theta_{\Omega}\left[\sqrt{3}\sqrt{1+\alpha_{\psi}-\alpha_{2}}\left(\sqrt{1-\alpha_{\psi}-\alpha_{1}}\sin\phi_{1}+\sqrt{1-\alpha_{\psi}+\alpha_{1}}\sin\phi_{4}\right)\right.\nonumber\\
       &\left.-3\sqrt{2}\sqrt{\alpha_{2}\left(1-\alpha_{\psi}-\alpha_{1}\right)}\sin\left(\phi_{1}+\phi_{3}\right)\right],\\
S_{13}= & -\frac{\sqrt{6}}{2}\sin^{2}\theta_{\Omega}\sqrt{\alpha_{2}\left(1+\alpha_{\psi}-\alpha_{2}\right)}\sin\phi_{3}.\label{eq:S13}
\end{align}
We also give the expressions for $S_i$ in terms of the helicity amplitudes $H_{1}$, $H_{2}$, $H_{3}$, and $H_{4}$ in Appendix~\ref{D.spin_correlations}, so that one can easily get the result for any parametrization scheme. 

Except for the $S_0$ term which corresponds to the cross section, the six non-zero polarization components listed above survive from the total 15 terms due to parity conservation. We will analyse this through the physical interpretation of these components as detailed in Appendix~\ref{B.Probabilistic}. In the parity-conserving $e^{+}e^{-} \rightarrow \Omega^{-}\bar{\Omega}^{+}$ process, a parity transformation leads to
\begin{align}
\hat{x}_{\Omega}\rightarrow-\hat{x}_{\Omega},\quad\hat{y}_{\Omega}\rightarrow\hat{y}_{\Omega},\quad\hat{z}_{\Omega}\rightarrow-\hat{z}_{\Omega}.
\end{align}
This transformation keeps the components $S_{T}^{y}\,(S_{3}/S_0)$, $S_{LL}\,(S_{4}/S_0)$, $S_{LT}^{x}\,(S_{5}/S_0)$, $S_{TT}^{xx}\,(S_{7}/S_0)$, $S_{LLT}^{y}\,(S_{11}/S_0)$, $S_{LTT}^{xy}\,(S_{13}/S_0)$, and $S_{TTT}^{yxx}\,(S_{15}/S_0)$ unchanged, demonstrating their compliance with parity conservation. Additionally, the limitation of helicity transitions to $\pm1$ leads to the absence of certain off-diagonal elements in the spin density matrix, as shown in Eq.~\eqref{eq:spin_density_matrix_Omega}. The component $S_{TTT}^{yxx}\,(S_{15}/S_0)$ is forbidden because it relates to the basis matrix $\Sigma_{15}$ as shown in Eq.~\eqref{eq:basis_matrix2}, which only have non-zero elements corresponding to helicity transitions beyond $\pm1$. Consequently, only the above six polarization components survive in this process.

\begin{figure}[ht]
\begin{centering}
\includegraphics[width=0.33\textwidth]{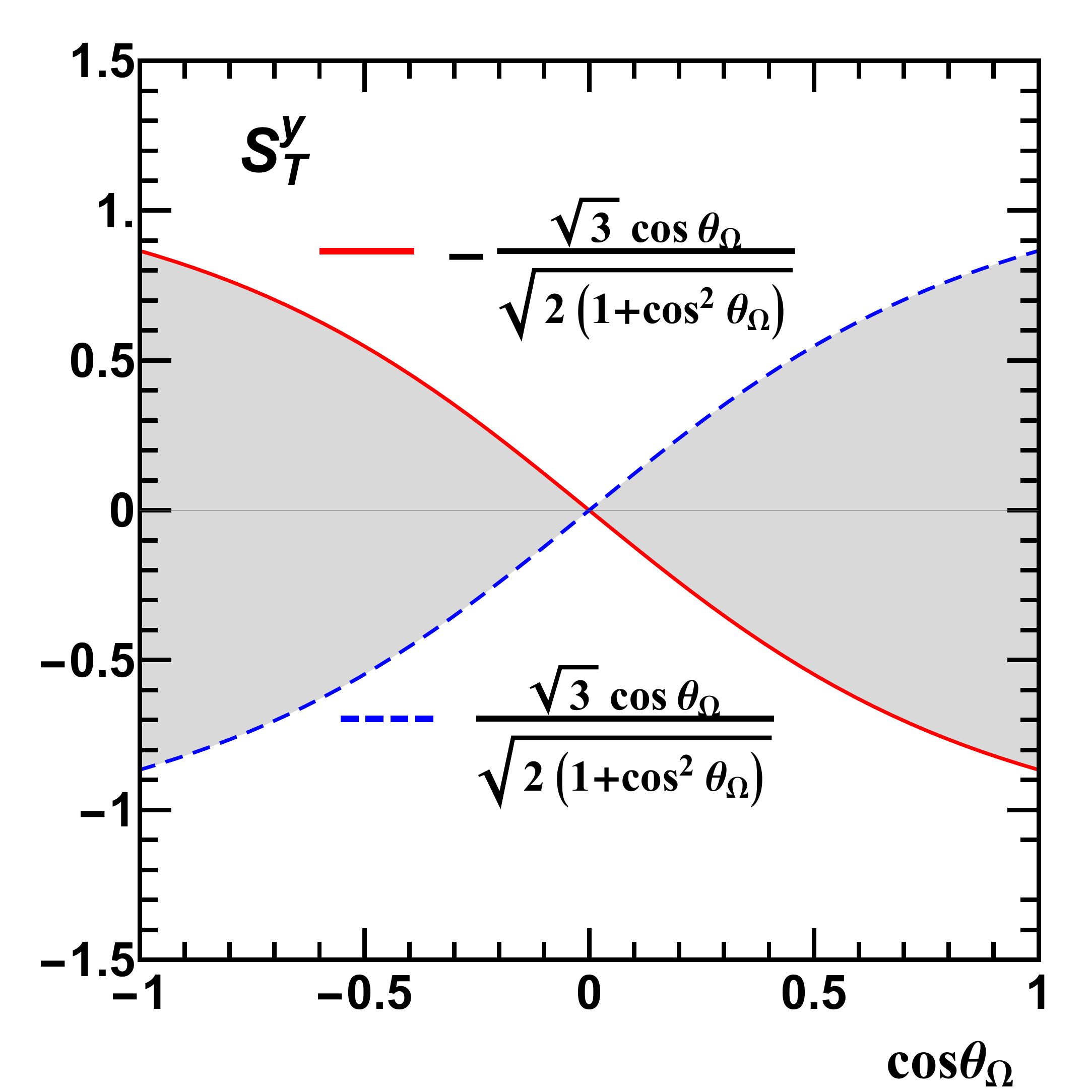}\includegraphics[width=0.33\textwidth]{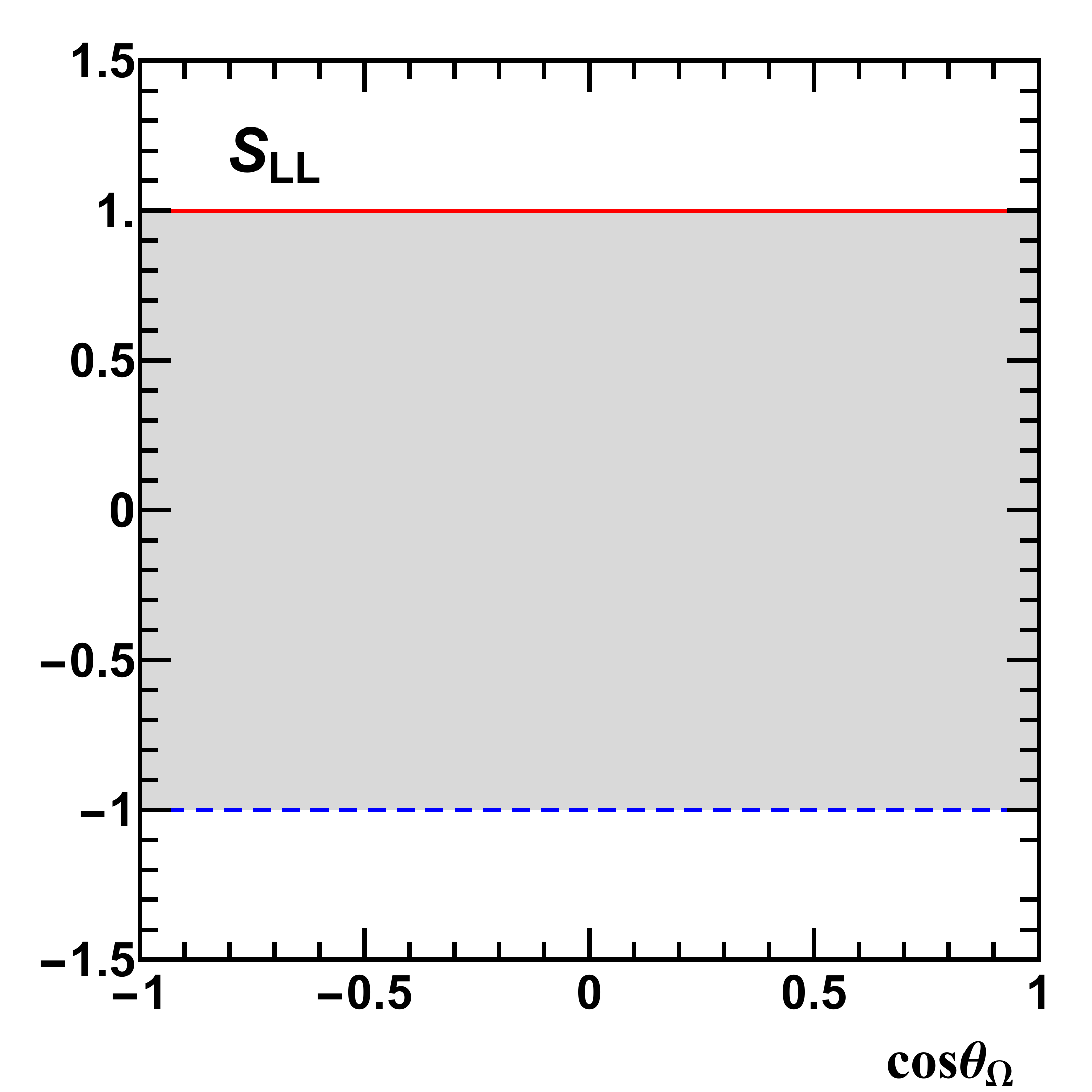}\includegraphics[width=0.33\textwidth]{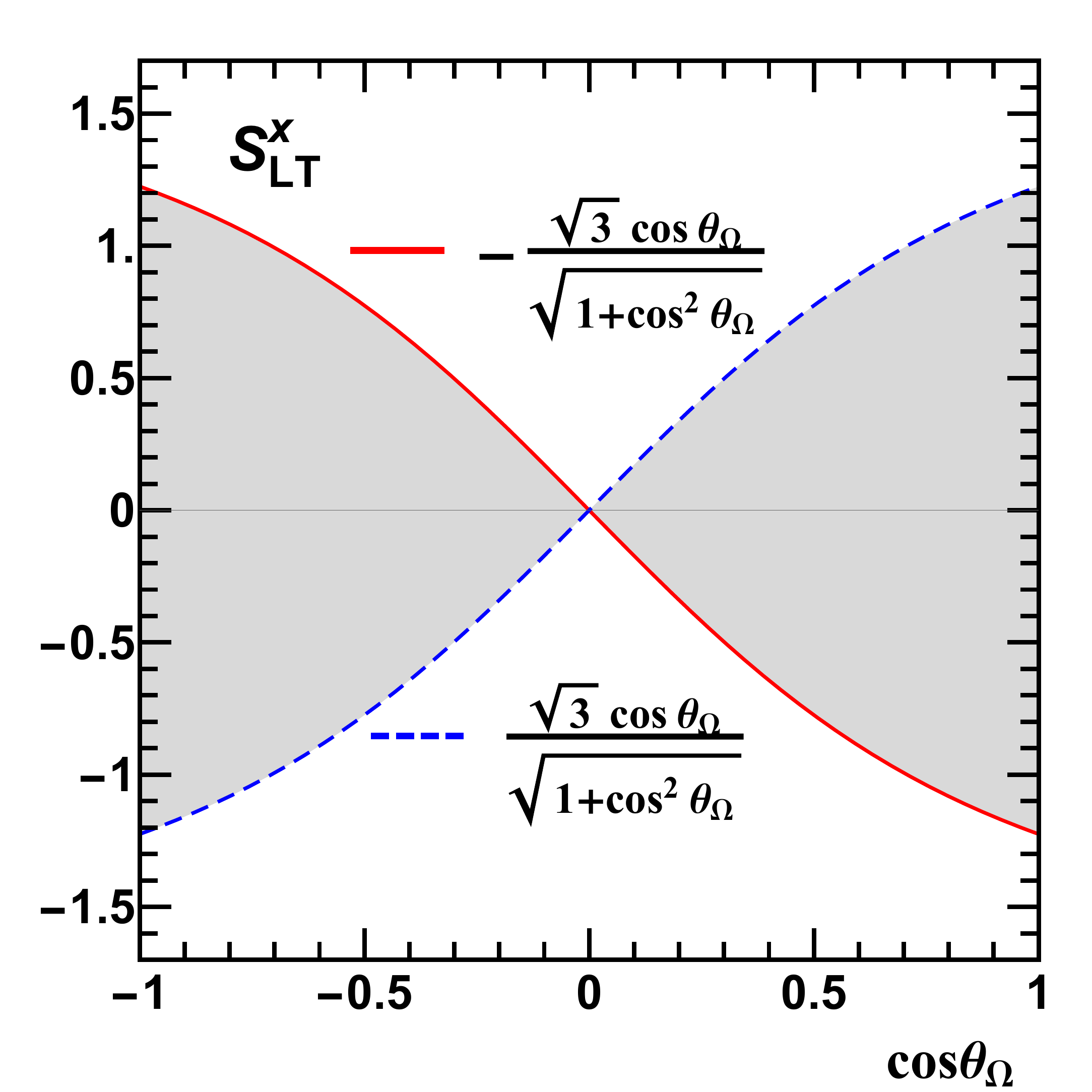}
\par\end{centering}
\begin{centering}
\includegraphics[width=0.33\textwidth]{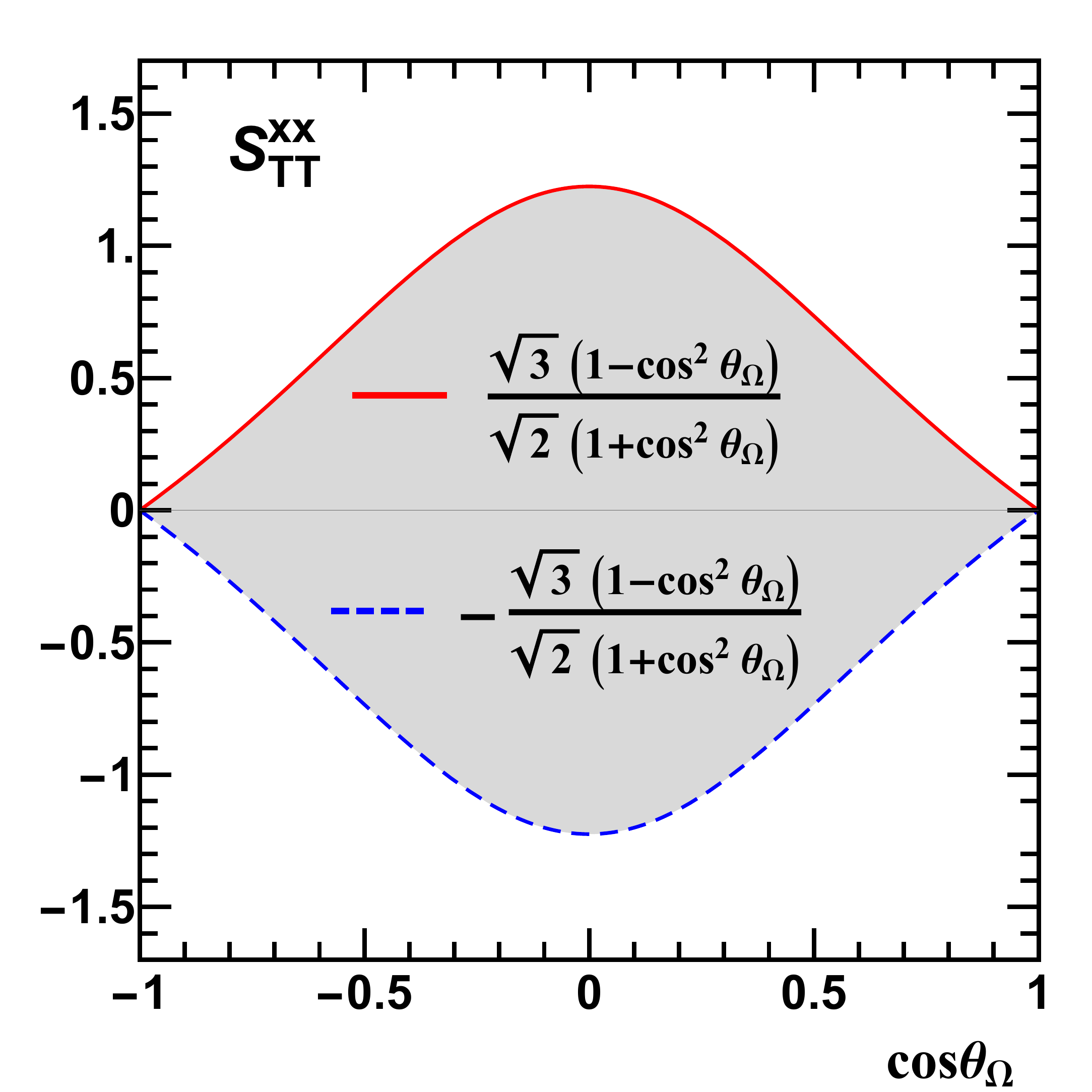}\includegraphics[width=0.33\textwidth]{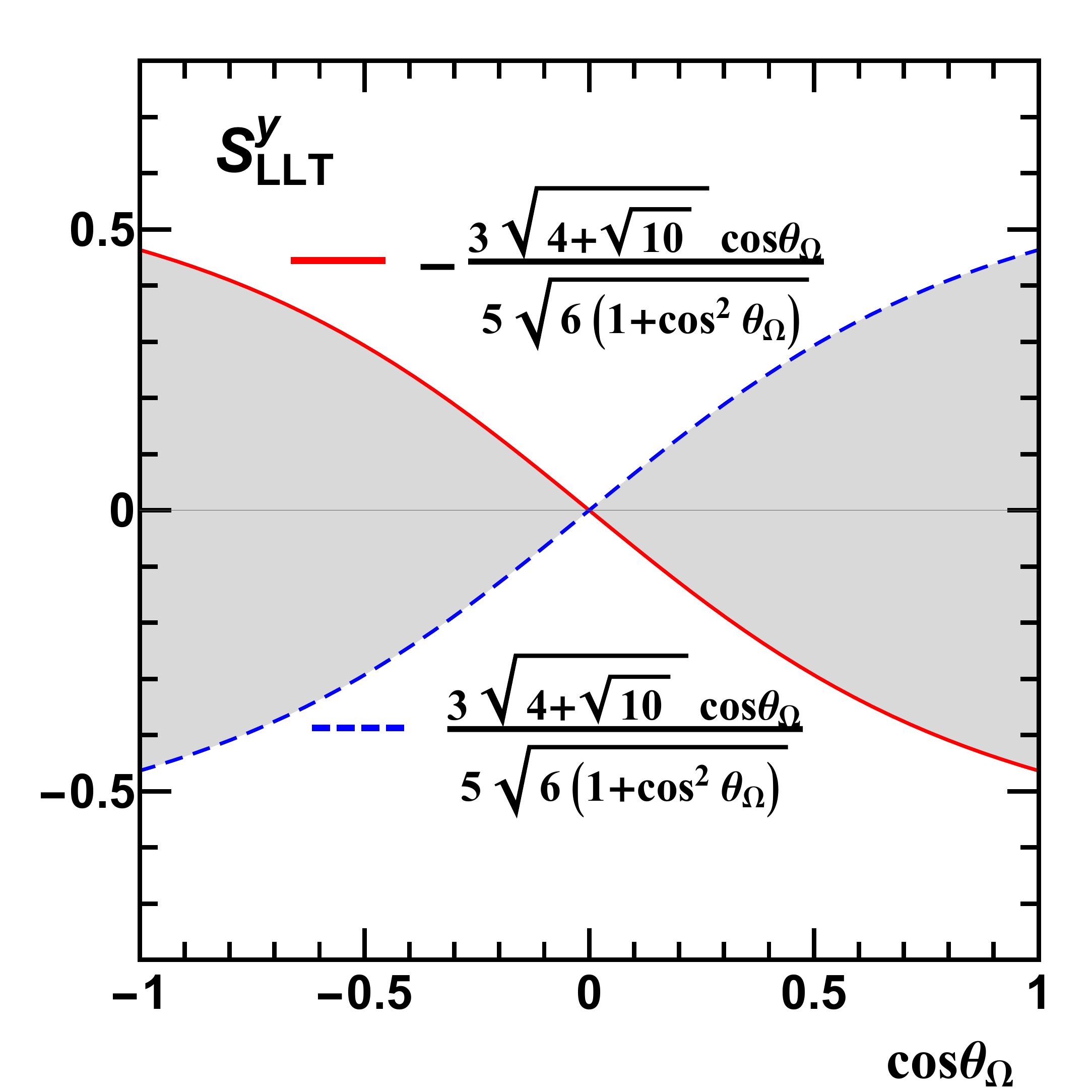}\includegraphics[width=0.33\textwidth]{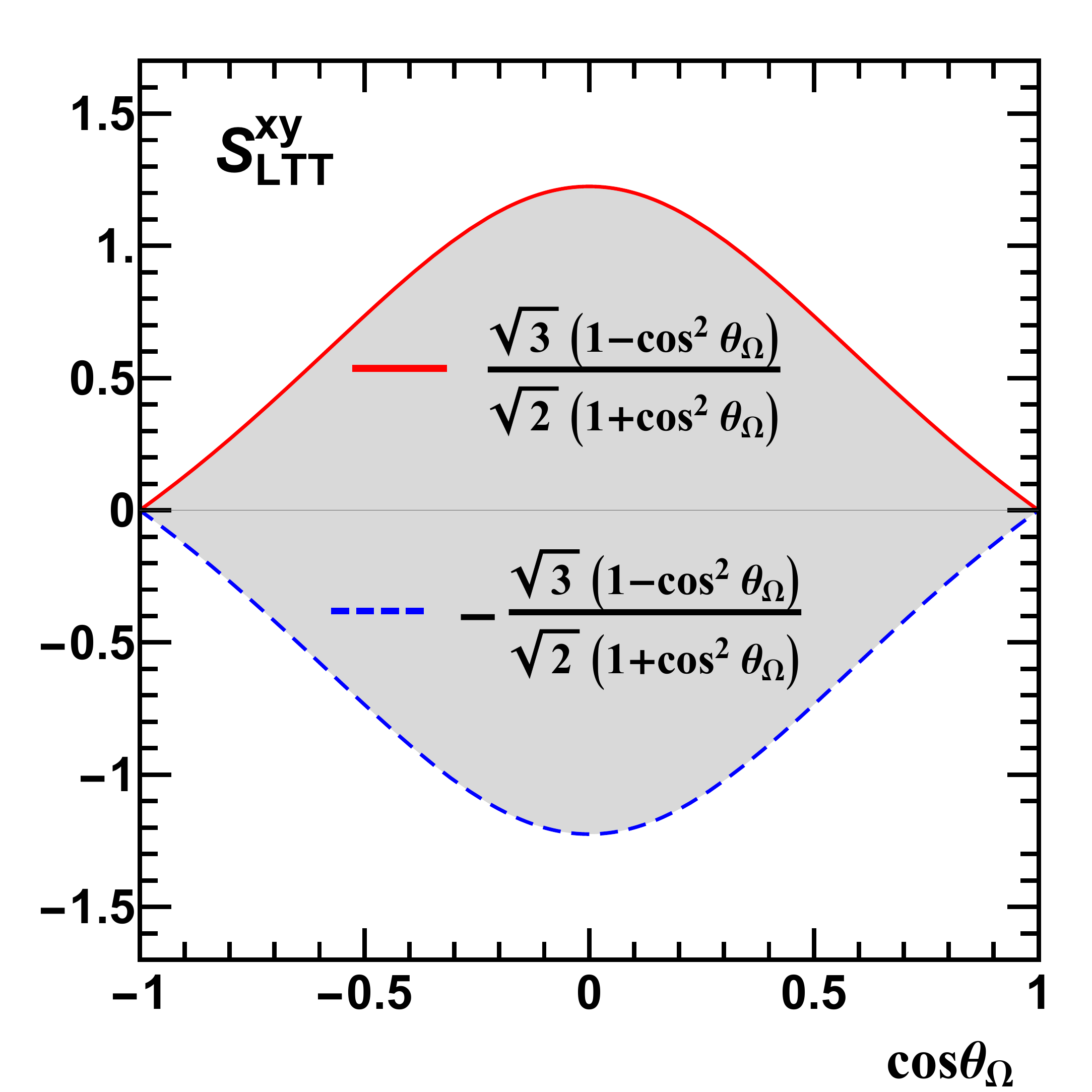}
\par\end{centering}
\caption{\label{fig:ranges} Boundaries of the spin components in single tag $\Omega^-$ process.}
\end{figure}

While the spin components have natural boundaries according to their physical interpretations, as listed in Eq.~\eqref{eq:spin_range}, the single-tag $\Omega^-$ measurement in $e^{+}e^{-}\rightarrow\Omega^{-}\bar{\Omega}^{+}$ process will also implement constraints on these polarization components. This effect can be analysed through applying the ranges for $\alpha_\psi,\alpha_1,\alpha_2$ as shown in Eqs.~\eqref{eq:range1}-\eqref{eq:range2} to the six non-zero spin components in this process. We illustrate the results in Fig.~\ref{fig:ranges}, with the red and blue curves denoting the boundary lines calculated from Eqs.~\eqref{eq:S0}-\eqref{eq:S13}. One can easily see that, these boundaries for spin components are narrower than those determined by their physical interpretations and 
are influenced by the production angle $\theta_{\Omega}$, except for $S_{LL}$. This finding offers new insights into the polarization dynamics of this process and deserves further research.

We also observe from the figure that the domains for all $\Omega^{-}$ polarization components include zero, which correspond to unpolarized states. Two sets of parameter values will lead to this specific situation, which are $\alpha_{1} =\alpha_{2} =0, \alpha_{\psi} =\pm 1$ and $\alpha_{1} =\alpha_{2} =0, \phi_4 = -\phi_1$. The comparison between these values and experimental results can reveal the polarization degree of the $\Omega^{-}$ particle, offering an intuitive understanding of the parameters.

\subsection{Double-tag $\Omega^{-}\bar{\Omega}^{+}$} \label{s.double}
In this subsection we analysis the spin states of the double-tag $\Omega^{-}\bar{\Omega}^{+}$. Similar to the single-tag case, we expand the density matrix in a complete set of orthogonal basis matrices, and the expanding coefficients correspond to different polarization states. In the double-tag case there are 256 polarization correlation coefficients, out of which 116 are nonzero.  These nonzero terms are linearly dependent, so we only list the terms necessary and the rest can be expressed by the linear combination of these terms. For reference, the original versions of these coefficients, expressed with variables $H_{1}$, $H_{2}$, $H_{3}$, and $H_{4}$, are available in Appendix~\ref{D.spin_correlations}.

For a spin-3/2 particle pair system, the general spin correlation is formulated as
\begin{align}
\rho_{B_{1}\bar{B}{2}} = \sum_{\mu=0}^{15}\sum_{\nu=0}^{15} S_{\mu,\nu} \Sigma_{\mu}^{B_{1}} \otimes \Sigma_{\nu}^{\bar{B}{2}},
\end{align}
where $S_{\mu,\nu}$  denotes the spin correlation coefficients. By comparing the above equation with the production density matrix for $\Omega^{-}\bar{\Omega}^{+}$ pairs as shown in Eq.~\eqref{eq:production density matrix}, we can obtain the  correlation coefficients. Charge conjugation invariance in the production process informs the following relationship,
\begin{align}
S_{\mu,\nu}\left(\theta_{\Omega}\right)= & S_{\nu,\mu}\left(\pi-\theta_{\Omega}\right).\label{eq:Smunu}
\end{align}
Given this, we only need to provide specific expressions for terms where $\mu \geq \nu$. For convenience, we introduce these $D$-type functions,
\begin{align}
D_{1}^{s} = & \sqrt{\left(1-\alpha_{\psi}-\alpha_{1}\right)\left(1+\alpha_{\psi}-\alpha_{2}\right)}\sin\phi_{1},\\
D_{1}^{c} = & \sqrt{\left(1-\alpha_{\psi}-\alpha_{1}\right)\left(1+\alpha_{\psi}-\alpha_{2}\right)}\cos\phi_{1},\\
D_{2}^{s} = & \sqrt{\alpha_{2}\left(1-\alpha_{\psi}-\alpha_{1}\right)}\sin\left(\phi_{1}+\phi_{3}\right),\\
D_{2}^{c} = & \sqrt{\alpha_{2}\left(1-\alpha_{\psi}-\alpha_{1}\right)}\cos\left(\phi_{1}+\phi_{3}\right),\\
D_{3}^{s} = & \sqrt{\alpha_{2}\left(1+\alpha_{\psi}-\alpha_{2}\right)}\sin\phi_{3},\\
D_{3}^{c} = & \sqrt{\alpha_{2}\left(1+\alpha_{\psi}-\alpha_{2}\right)}\cos\phi_{3},\\
D_{4}^{s} = & \sqrt{\left(1-\alpha_{\psi}+\alpha_{1}\right)\left(1+\alpha_{\psi}-\alpha_{2}\right)}\sin\phi_{4},\\
D_{4}^{c} = & \sqrt{\left(1-\alpha_{\psi}+\alpha_{1}\right)\left(1+\alpha_{\psi}-\alpha_{2}\right)}\cos\phi_{4},\\
D_{5}^{s} = & \sqrt{\left(1-\alpha_{\psi}-\alpha_{1}\right)\left(1-\alpha_{\psi}+\alpha_{1}\right)}\sin\left(\phi_{1}-\phi_{4}\right),\\
D_{5}^{c} = & \sqrt{\left(1-\alpha_{\psi}-\alpha_{1}\right)\left(1-\alpha_{\psi}+\alpha_{1}\right)}\cos\left(\phi_{1}-\phi_{4}\right),\\
D_{6}^{s} = & \sqrt{\alpha_{2}\left(1-\alpha_{\psi}+\alpha_{1}\right)}\sin\left(\phi_{3}+\phi_{4}\right),\\
D_{6}^{c} = & \sqrt{\alpha_{2}\left(1-\alpha_{\psi}+\alpha_{1}\right)}\cos\left(\phi_{3}+\phi_{4}\right).
\end{align}

We now outline the explicit expressions for the diagonal coefficients.  We identify the 13 independent terms among the total 16 diagonal coefficients,
\begin{align}
S_{0,0}= & 1+\alpha_{\psi}\cos^{2}\theta_{\Omega},\label{Smunu1}\\
S_{1,1}= & \frac{1}{4}\left(4-\alpha_{\psi}+2\alpha_{1}-2\alpha_{2}\right)-\frac{1}{4}\left(1-4\alpha_{\psi}+2\alpha_{1}+2\alpha_{2}\right)\cos^{2}\theta_{\Omega}\\
S_{2,2}= & \frac{1}{8}\left(5+\alpha_{\psi}-2\alpha_{1}+\alpha_{2}+3D_{5}^{c}\right)\sin^{2}\theta_{\Omega}+\frac{\sqrt{6}}{4}D_{3}^{c}\left(3+\cos2\theta_{\Omega}\right),\\
S_{3,3}= & \frac{1}{8}\left(1+5\alpha_{\psi}+2\alpha_{1}+\alpha_{2}-3D_{5}^{c}\right)\sin^{2}\theta_{\Omega}-\frac{\sqrt{6}}{4}D_{3}^{c}\left(3+\cos2\theta_{\Omega}\right),\\
S_{4,4}= & -\left(\alpha_{\psi}-\alpha_{2}\right)-\left(1-\alpha_{2}\right)\cos^{2}\theta_{\Omega},\\
S_{5,5}= & \frac{3}{2}\left(1+\alpha_{\psi}-\alpha_{2}+D_{5}^{c}\right)\sin^{2}\theta_{\Omega},\\
S_{6,6}= & \frac{3}{2}\left(1+\alpha_{\psi}-\alpha_{2}-D_{5}^{c}\right)\sin^{2}\theta_{\Omega},\\
S_{7,7}= & \frac{3}{2}D_{5}^{c}\sin^{2}\theta_{\Omega}+\frac{3}{4}\left(1+\alpha_{\psi}-\alpha_{2}\right)\left(3+\cos2\theta_{\Omega}\right),\\
S_{9,9}= & \frac{9}{100}\left(1-4\alpha_{\psi}-2\alpha_{1}-3\alpha_{2}\right)-\frac{9}{100}\left(4-\alpha_{\psi}-2\alpha_{1}+3\alpha_{2}\right)\cos^{2}\theta_{\Omega},\\
S_{10,10}= & \frac{3}{100}\left(5-\alpha_{\psi}-3\alpha_{1}+4\alpha_{2}+2D_{5}^{c}\right)\sin^{2}\theta_{\Omega}-\frac{3\sqrt{6}}{50}D_{3}^{c}\left(3+\cos2\theta_{\Omega}\right),\\
S_{11,11}= & -\frac{3}{100}\left(1-5\alpha_{\psi}-3\alpha_{1}-4\alpha_{2}+2D_{5}^{c}\right)\sin^{2}\theta_{\Omega} +\frac{3\sqrt{6}}{50}D_{3}^{c}\left(3+\cos2\theta_{\Omega}\right),\\
S_{12,12}= & \frac{3}{2}D_{5}^{c}\sin^{2}\theta_{\Omega}-\frac{3}{4}\left(1+\alpha_{\psi}-\alpha_{2}\right)\left(3+\cos2\theta_{\Omega}\right),\\
S_{14,14}= & \frac{9}{4}\left(1-\alpha_{\psi}+\alpha_{1}\right)\sin^{2}\theta_{\Omega}.\label{Smunu2}
\end{align}
The remaining three dependent diagonal coefficients are
\begin{align}
\left\{ S_{8,8},S_{13,13},S_{15,15}\right\}  = \left\{ -S_{7,7},-S_{12,12},-S_{14,14}\right\} ,\label{eq:independent1}
\end{align}
where each coefficient on the left side corresponds directly to its counterpart on the right side, exemplified by $S_{8,8} = -S_{7,7}$.

For the off-diagonal coefficients, we present a subset of $S_{\mu,\nu}$ with $\mu>\nu$, totaling 50 non-zero coefficients. These coefficients are classified according to the number of $D$-type functions contained in their expressions. There are six coefficients without $D$-type functions, with the following three selected as the independent ones,
\begin{align}
S_{4,0} & = \frac{1}{2}\left(\alpha_{1}-\alpha_{2}\right)-\frac{1}{2}\left(\alpha_{1}+\alpha_{2}\right)\cos^{2}\theta_{\Omega},\label{Smunu0_1}\\
S_{9,1} & =  -\frac{3}{40}\left(4+4\alpha_{\psi}-3\alpha_{1}-7\alpha_{2}\right)-\frac{3}{40}\left(4+4\alpha_{\psi}+3\alpha_{1}-7\alpha_{2}\right)\cos^{2}\theta_{\Omega},\\
S_{14,2} & =  \frac{3}{4}\left(1+\alpha_{\psi}-\alpha_{2}\right)\sin^{2}\theta_{\Omega},\label{Smunu0_2}
\end{align}
and the remaining three are represented as
\begin{align}
\left\{ S_{14,10},S_{15,3},S_{15,11}\right\}  = \left\{ -\frac{3}{5}S_{14,2},-S_{14,2},\frac{3}{5}S_{14,2}\right\} .\label{eq:independent2}
\end{align}

For case where the $D$-type functions appear once, there are a total of 18 coefficients. The following seven are chosen as independent ones,
\begin{align}
S_{7,0}  = & \frac{\sqrt{6}}{2}D_{3}^{c}\sin^{2}\theta_{\Omega},\label{Smunu1_1}\\
S_{13,0}  = & -\frac{\sqrt{6}}{2}D_{3}^{s}\sin^{2}\theta_{\Omega},\\
S_{13,7}  = & \frac{3}{2}D_{5}^{s}\sin^{2}\theta_{\Omega},\\
S_{7,5} = & \frac{3\sqrt{2}}{4}D_{6}^{c}\sin2\theta_{\Omega},\\
S_{13,5}  = & -\frac{3\sqrt{2}}{4}D_{6}^{s}\sin2\theta_{\Omega},\\
S_{15,7}  = & -\frac{3\sqrt{3}}{4}D_{4}^{s}\sin2\theta_{\Omega},\\
S_{14,12}  = & \frac{3\sqrt{3}}{4}D_{4}^{c}\sin2\theta_{\Omega},\label{Smunu1_2}
\end{align}
and the remaining 11 coefficients can be expressed as,
\begin{align}
\left\{ S_{12,1},S_{7,4},S_{12,9}\right\}  = & \left\{ \frac{1}{2}S_{7,0},-S_{7,0},-\frac{9}{10}S_{7,0}\right\}, \label{eq:independent3}\\
\left\{ S_{8,1},S_{13,4},S_{9,8}\right\}  = & \left\{ \frac{1}{2}S_{13,0},-S_{13,0},-\frac{9}{10}S_{13,0}\right\}, \\
\left\{ S_{8,6},S_{12,6},S_{12,8},S_{14,8},S_{15,13}\right\}  = & \left\{ -S_{7,5},S_{13,5},S_{13,7},S_{15,7},-S_{14,12}\right\}. \label{eq:independent4}
\end{align}

When the $D$-type functions appear twice,  there are 18 coefficients, with these 12 listed here as independent ones,
\begin{align}
S_{5,0} = & -\frac{\sqrt{3}}{4}\left(D_{1}^{c}-D_{4}^{c}\right)\sin2\theta_{\Omega},\label{Smunu2_1}\\
S_{6,1} = & -\frac{\sqrt{3}}{8}\left(D_{1}^{s}+3D_{4}^{s}\right)\sin2\theta_{\Omega},\\
S_{6,2} = & \frac{3}{4}D_{5}^{s}\sin^{2}\theta_{\Omega}-\frac{\sqrt{6}}{4}D_{3}^{s}\left(3+\cos2\theta_{\Omega}\right),\\
S_{8,2} = & -\frac{1}{8}\left(2\sqrt{3}D_{1}^{s}+3\sqrt{2}D_{6}^{s}\right)\sin2\theta_{\Omega},\\
S_{10,2} = & \frac{3}{20}\left(2\alpha_{\psi}+\alpha_{1}-3\alpha_{2}+D_{5}^{c}\right)\sin^{2}\theta_{\Omega}
             -\frac{\sqrt{6}}{40}D_{3}^{c}\left(3+\cos2\theta_{\Omega}\right),\\
S_{12,2} = & -\frac{1}{8}\left(2\sqrt{3}D_{1}^{c}-3\sqrt{2}D_{6}^{c}\right)\sin2\theta_{\Omega},\\
S_{11,3} = & \frac{3}{20}\left(2-\alpha_{1}-3\alpha_{2}
              -D_{5}^{c}\right)\sin^{2}\theta_{\Omega}+\frac{\sqrt{6}}{40}D_{3}^{c}\left(3+\cos2\theta_{\Omega}\right)\\
S_{5,4} = & \frac{\sqrt{3}}{4}\left(D_{1}^{c}+D_{4}^{c}\right)\sin2\theta_{\Omega},\\
S_{11,5} = & \frac{3}{10}D_{5}^{s}\sin^{2}\theta_{\Omega}+\frac{3\sqrt{6}}{20}D_{3}^{s}\left(3+\cos2\theta_{\Omega}\right),\\
S_{9,6} = & -\frac{3\sqrt{3}}{40}\left(3D_{1}^{s}-D_{4}^{s}\right)\sin2\theta_{\Omega},\\
S_{11,7} = & -\frac{3}{20}\left(\sqrt{3}D_{1}^{s}-\sqrt{2}D_{6}^{s}\right)\sin2\theta_{\Omega},\\
S_{12,10} = & \frac{3}{20}\left(\sqrt{3}D_{1}^{c}+\sqrt{2}D_{6}^{c}\right)\sin2\theta_{\Omega},\label{Smunu2_2}
\end{align}
and the rest six ones are written as,
\begin{align}
\left\{ S_{5,3},S_{10,6}, S_{7,3}\right\}  = & \left\{ S_{6,2},S_{11,5},S_{8,2}\right\} ,\label{eq:independent5}\\
\left\{S_{13,3},S_{10,8},S_{13,11}\right\}  = & \left\{ -S_{12,2},S_{11,7},-S_{12,10}\right\} .\label{eq:independent6}
\end{align}

Finally, when the $D$-type functions appear three times, there are eight coefficients, all of which are independent,
\begin{align}
S_{3,0} = & -\frac{1}{8}\left[2\sqrt{2}D_{2}^{s}+\sqrt{3}\left(D_{1}^{s}+D_{4}^{s}\right)\right]\sin2\theta_{\Omega},\label{Smunu3_1}\\
S_{11,0} = & \frac{1}{20}\left[3\sqrt{2}D_{2}^{s}-\sqrt{3}\left(D_{1}^{s}+D_{4}^{s}\right)\right]\sin2\theta_{\Omega},\\
S_{2,1} = & \frac{1}{16}\left[2\sqrt{2}D_{2}^{c}-\sqrt{3}\left(D_{1}^{c}-3D_{4}^{c}\right)\right]\sin2\theta_{\Omega},\\
S_{10,1} = & -\frac{1}{40}\left[3\sqrt{2}D_{2}^{c}+\sqrt{3}\left(D_{1}^{c}-3D_{4}^{c}\right)\right]\sin2\theta_{\Omega},\\
S_{9,2} = & \frac{3}{80}\left[6\sqrt{2}D_{2}^{c}-\sqrt{3}\left(3D_{1}^{c}+D_{4}^{c}\right)\right]\sin2\theta_{\Omega},\\
S_{4,3} = & -\frac{1}{8}\left[2\sqrt{2}D_{2}^{s}+\sqrt{3}\left(D_{1}^{s}-D_{4}^{s}\right)\right]\sin2\theta_{\Omega},\\
S_{11,4} = & -\frac{1}{20}\left[3\sqrt{2}D_{2}^{s}-\sqrt{3}\left(D_{1}^{s}-D_{4}^{s}\right)\right]\sin2\theta_{\Omega},\\
S_{10,9} = & \frac{3}{200}\left[9\sqrt{2}D_{2}^{c}+\sqrt{3}\left(3D_{1}^{c}+D_{4}^{c}\right)\right]\sin2\theta_{\Omega}.\label{Smunu3_2}
\end{align}

We have detailed the 66 polarization correlation coefficients where $\mu \geq \nu$. The correlation coefficients with $\mu < \nu$ can be derived using the relation $S_{\mu,\nu}(\theta_{\Omega}) = S_{\nu,\mu}(\pi - \theta_{\Omega})$. Among these 50 coefficients, 22 show symmetry in the exchange of $\mu$ and $\nu$, and the remaining 28 coefficients are anti-symmetric. The symmetric ones are
\begin{align}
\left\{ \begin{array}{c}
S_{0,4},S_{0,7},S_{0,13},S_{1,8},S_{1,9},S_{1,12}\\
S_{2,6},S_{2,10},S_{2,14},S_{3,5},S_{3,11},S_{3,15}\\
S_{4,7},S_{4,13},S_{5,11},S_{6,10},S_{7,13}\\
S_{8,9},S_{8,12},S_{9,12},S_{10,14},S_{11,15}
\end{array}\right\}
= \left\{ \begin{array}{c}
S_{4,0},S_{7,0},S_{13,0},S_{8,1},S_{9,1},S_{12,1}\\
S_{6,2},S_{10,2},S_{14,2},S_{5,3},S_{11,3},S_{15,3}\\
S_{7,4},S_{13,4},S_{11,5},S_{10,6},S_{13,7}\\
S_{9,8},S_{12,8},S_{12,9},S_{14,10},S_{15,11}
\end{array}\right\},\label{eq:independent7}
\end{align}
and the coefficients that exhibit anti-symmetry are
\begin{align}
\left\{ \begin{array}{c}
S_{0,3},S_{0,5},S_{0,11},S_{1,2},S_{1,6},S_{1,10}\\
S_{2,8},S_{2,9},S_{2,12},S_{3,4},S_{3,7},S_{3,13}\\
S_{4,5},S_{4,11}S_{5,7},S_{5,13},S_{6,8},S_{6,9}\\
S_{6,12},S_{7,11},S_{7,15},S_{8,10},S_{8,14}\\
S_{9,10},S_{10,12},S_{11,13},S_{12,14},S_{13,15}
\end{array}\right\}
= -\left\{ \begin{array}{c}
S_{3,0},S_{5,0},S_{11,0},S_{2,1},S_{6,1},S_{10,1}\\
S_{8,2},S_{9,2},S_{12,2},S_{4,3},S_{7,3},S_{13,3}\\
S_{5,4},S_{11,4}S_{7,5},S_{13,5},S_{8,6},S_{9,6}\\
S_{12,6},S_{11,7},S_{15,7},S_{10,8},S_{14,8}\\
S_{10,9},S_{12,10},S_{13,11}S_{14,12},S_{15,13}
\end{array}\right\} .\label{eq:independent8}
\end{align}
To explore the reasons behind the symmetry and anti-symmetry in these coefficients, we consider CP conservation in the $e^{+}e^{-} \rightarrow \Omega^{-}\bar{\Omega}^{+}$ process. Due to Eq.~\eqref{eq:Omega_Omegabar_axis} and  Fig.~\ref{fig:Omega},  a CP transformation leads to modifications in the coordinate systems of $\Omega^{-}$ and $\bar{\Omega}^{+}$,
\begin{align}
\hat{x}_{\Omega} & \rightarrow-\hat{x}_{\Omega},\quad\hat{y}_{\Omega}\rightarrow-\hat{y}_{\Omega}\quad\hat{z}_{\Omega}\rightarrow\hat{z}_{\Omega},\\
\hat{x}_{\bar{\Omega}} & \rightarrow-\hat{x}_{\bar{\Omega}},\quad\hat{y}_{\bar{\Omega}}\rightarrow-\hat{y}_{\bar{\Omega}}\quad\hat{z}_{\bar{\Omega}}\rightarrow\hat{z}_{\bar{\Omega}},
\end{align}
where the transverse coordinates ($x$ and $y$) of both $\Omega^{-}$ and $\bar{\Omega}^{+}$ invert, and the longitudinal coordinates ($z$) remain unchanged.  Consequently, polarization coefficients involving an even number of transverse indices remain the same. For example, the coefficient $S_{1,8}$ (correspond to $S_{{L},{TT}^{xy}}$), containing two transverse indices, remains unchanged.  In contrast, coefficients with an odd number of transverse indices, such as $S_{1,2}$ (correspond to $S_{{L},{T}^{x}}$), undergo a sign change.

In our analysis of single-tag $\Omega^-$ polarization, we identify two solution sets indicating an unpolarized state of $\Omega^-$. For the double-tag $\Omega^-\bar{\Omega}^+$ system, there is no specific solution that completely zeros all polarization correlation coefficients, which corresponds to unpolarized state of the system. However, we find a  special solution set that zeros out all off-diagonal polarization correlation coefficients, significantly reducing polarization correlation in the $\Omega^-\bar{\Omega}^+$ system.  This particular solution is,
\begin{align}
\alpha_{1} = 0,
\quad
\alpha_{2} = 0,
\quad
\alpha_{\psi} = -1,
\quad
\phi_{4} = -\phi_{1}.
\end{align}
Possible discrepancy between these theoretical ideal values and experimental measurements would offer insight into the extent of polarization correlation between $\Omega^{-}$ and $\bar{\Omega}^{+}$.

\begin{figure}[ht]
\begin{centering}
\includegraphics[width=0.33\textwidth]{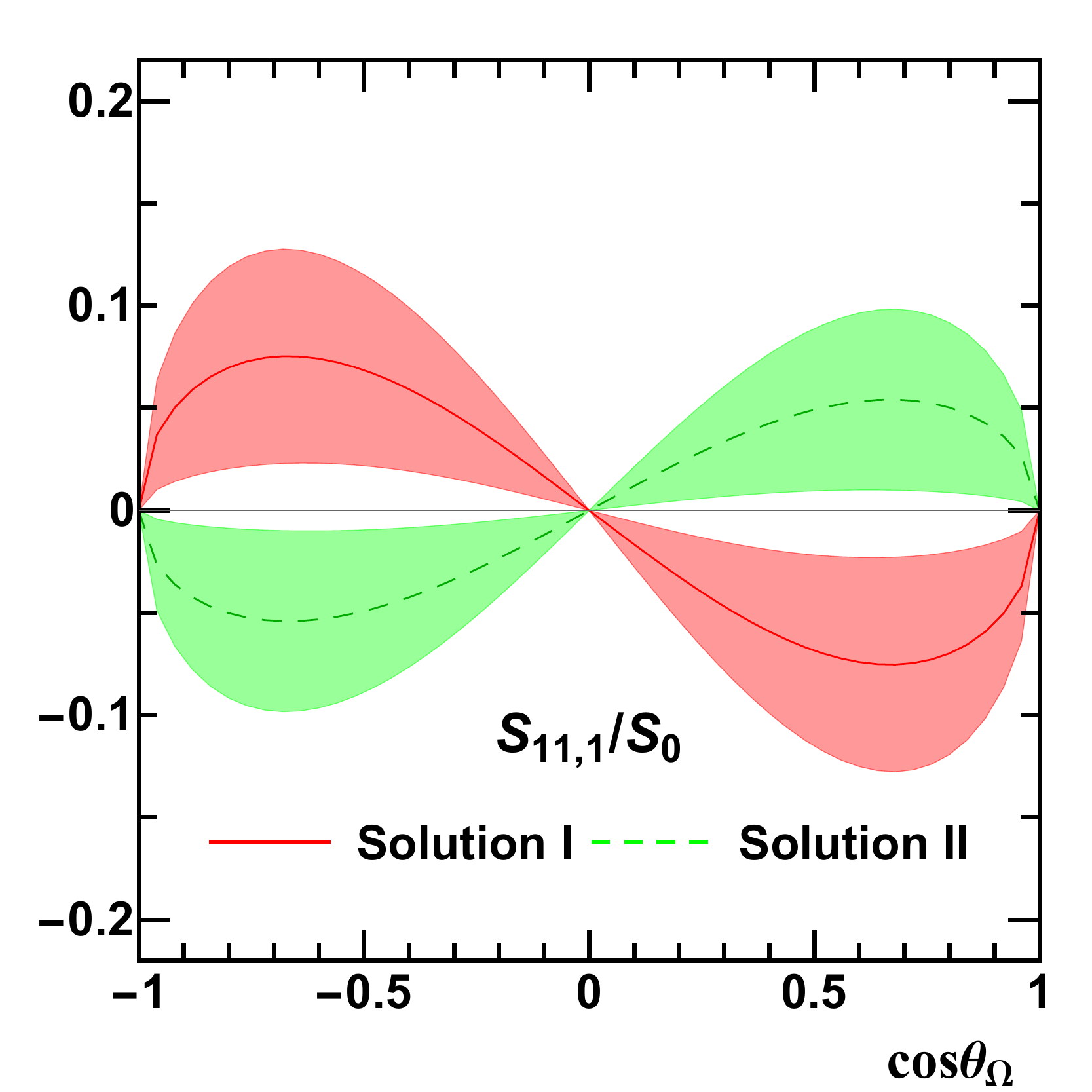}\includegraphics[width=0.33\textwidth]{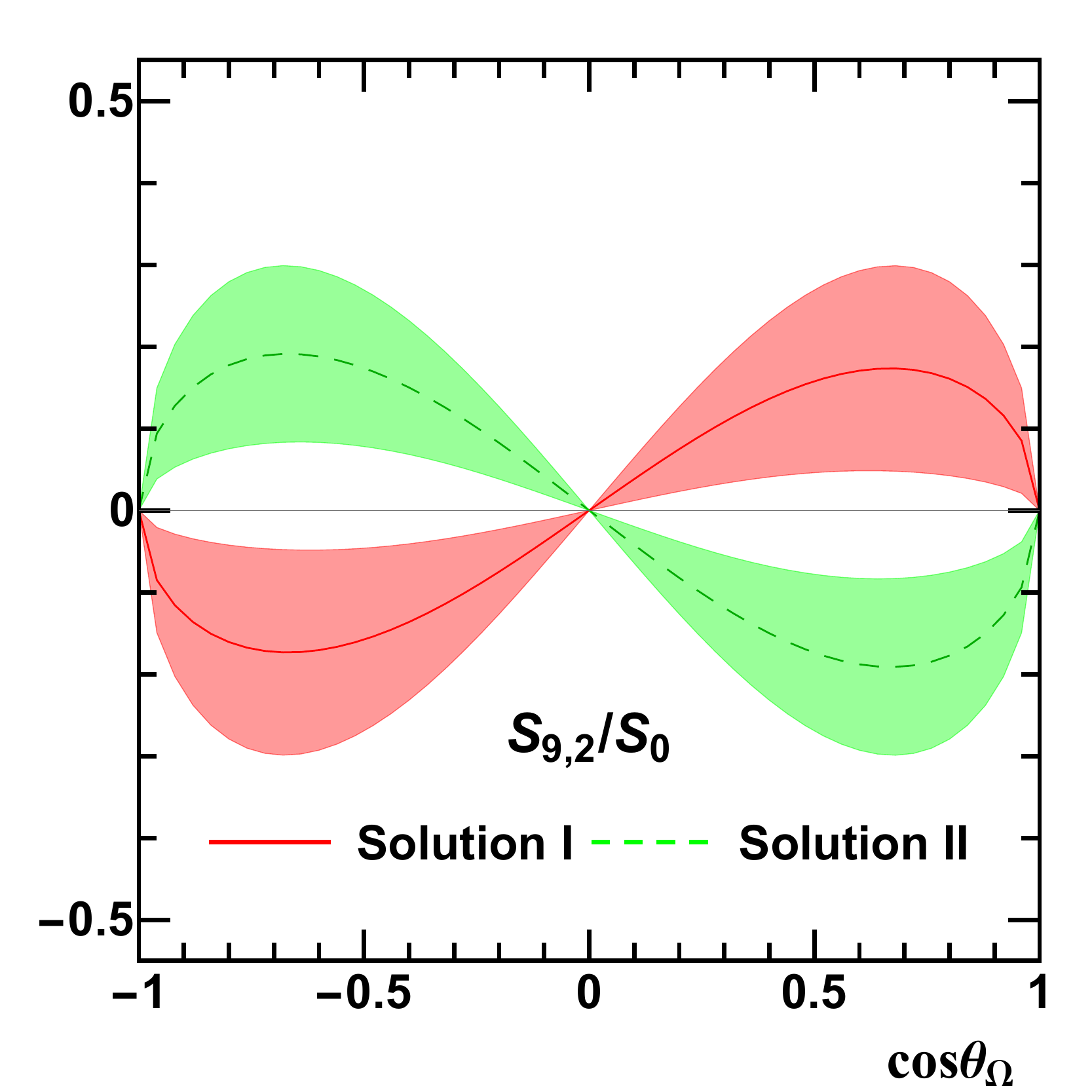}\includegraphics[width=0.33\textwidth]{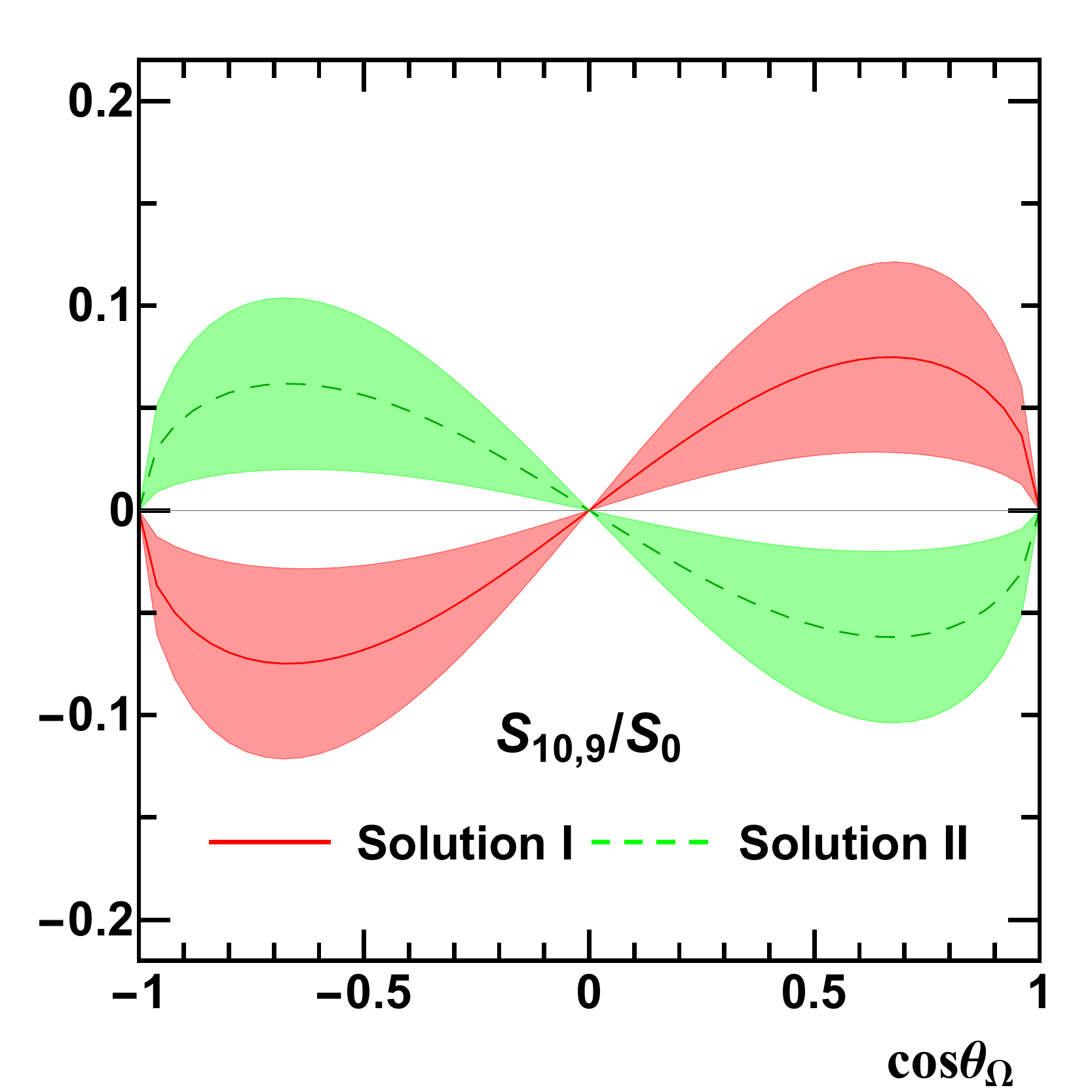}
\par\end{centering}
\caption{\label{fig:correlation components} Parts of the polarization correlation coefficients in the double-tag $\Omega^-\bar{\Omega}^+$ process with the two solutions measured in single-tag $\Omega^-$ process as inputs.}
\end{figure}

In single-tag $\Omega^{-}$ polarization measurements, there exists two sets of solutions correspond to the same polarization state, measured by BESIII~\cite{BESIII:2020lkm}. While in a double tag  $\Omega^{-}\bar{\Omega}^{+}$ measurement we claim that this uncertainty can be removed. To show this we insert the two solutions listed in Table \ref{tab:parametrization} into the polarization correlation coefficients we obtained in this subsection, and the results demonstrate that these two solutions can be separated clearly. We take $S_{11,1}$, $S_{9,2}$, and $S_{10,9}$ as examples in Fig.~\ref{fig:correlation components}. The conclusion is that it would be much easier to eliminate non-physical solutions in the double tag  $\Omega^{-}\bar{\Omega}^{+}$ measurement.

\section{decay chains} \label{s.decay}
Particles polarization states are typically inferred from decay processes, which manifest the polarization of parent particles in the angular distribution of their decay products. There are several approaches to describe particle decay. For instance, the polarization transfer matrix $a_{\mu\nu}$~\cite{Perotti:2018wxm} and the Lee-Yang formula~\cite{Lee:1957qs} are two common approaches for spin-1/2 particles. In this section, we present three methods to describe the decay expressions of spin-1/2 and spin-3/2 particles. The first method employs polarization transfer matrices, represented as $a_{\mu\nu}$ for spin-1/2 and $b_{\mu\nu}$ for spin-3/2 particles, within the helicity formalism.  The second approach applies the Lee-Yang formula for spin-1/2 particles and its adapted version for spin-3/2 particles. The third method is a new formulation we have developed. While these approaches are equivalent, each provides distinct insights into the decay processes.  Additionally, we discuss the discrepancies in understanding the decay parameters of the $\Omega$ particle among existing experiments.

Due to their computational simplicity, we begin with the use of polarization transfer matrices within the helicity formalism to describe decay expressions. For the decay of spin-1/2 particles, e.g., the $\Lambda \rightarrow p\pi^{-}$ process, the spin density matrix of the proton is represented as~\cite{Perotti:2018wxm}
\begin{align}
\rho_{1/2}^{p} =\sum_{\mu=0}^{3}\sum_{\nu=0}^{3}S_{\mu}a_{\mu\nu}\sigma_{\nu},
\end{align}
where $S_{\mu}$ denotes polarization of the initial $\Lambda$ particle, and $a_{\mu\nu}$ is the polarization transfer matrix, illustrating the polarization transition from the parent $\Lambda$ to the daughter proton. The polarization transfer matrix is expressed as
\begin{align}
a_{\mu\nu} =& \frac{1}{4\pi}\sum_{\lambda,\lambda^{\prime}=-1/2}^{1/2}\sum_{\kappa,\kappa^{\prime}=-1/2}^{1/2}B_{\lambda}B_{\lambda^{\prime}}^{*}\left(\sigma_{\mu}\right)^{\kappa,\kappa^{\prime}}\left(\sigma_{\nu}\right)^{\lambda^{\prime}\lambda}\mathcal{D}_{\kappa,\lambda}^{1/2*}\left(\Omega\right)\mathcal{D}_{\kappa^{\prime},\lambda^{\prime}}^{1/2}\left(\Omega\right),
\end{align}
where $\mathcal{D}_{\kappa,\lambda}^{J}\left(\Omega\right)=\mathcal{D}_{\kappa,\lambda}^{J}\left(0,\theta,\phi\right)$ represents the Wigner $D$-matrix, and $B_{\lambda}$ is the helicity amplitude for the  $\Lambda \rightarrow p\pi^{-}$ process with $\lambda$ being the helicity of the proton. The relation between these helicity amplitudes and the canonical amplitudes $A_L$ is given by
\begin{align}
B_{\lambda} =\sum_{L}\left(\frac{2L+1}{2J+1}\right)^{1/2}\left\langle L,0;S,\lambda|J,\lambda\right\rangle A_{L},\label{eq:canonical amplitudes}
\end{align}
where $\left\langle L,0;S,\lambda|J,\lambda\right\rangle $ denotes the Clebsch-Gordan coefficients, which involve spin  of the parent particle ($J$), the spin of the daughter particle ($S$), and the orbital angular momentum ($L$). For the case where $J=1/2$ and $S=1/2$, we obtain,
\begin{align}
B_{-1/2} = & \frac{\sqrt{2}}{2}\left(A_{S}+A_{P}\right),\label{eq:Bm1h_1/2}\\
B_{1/2} = & \frac{\sqrt{2}}{2}\left(A_{S}-A_{P}\right).\label{eq:B1h_1/2}
\end{align}
These helicity amplitudes  encompass both parity-conserving and parity-violating effects. For parity conservation processes, there is a relationship
\begin{align}
B_{\lambda} =\eta_{1}\eta_{2}\eta\left(-1\right)^{J-S}B_{-\lambda},\label{eq:parity}
\end{align}
where $\eta$, $\eta_{1}$, and $\eta_{2}$ denote the parities of the parent particle $\Lambda$ and its decay products, proton and $\pi^{-}$, respectively. By comparing Eqs.~\eqref{eq:Bm1h_1/2}-\eqref{eq:B1h_1/2} with Eq.~\eqref{eq:parity}, we identify that the $P$-wave term $(A_{P})$  aligns with parity conservation, while the $S$-wave term $(A_{S})$ indicates parity-violating transition. Using the Lee-Yang parametrization scheme with the normalization constraint $A_{S}^{2} + A_{P}^{2} = 1$, these amplitudes are parameterized as,
\begin{align}
\alpha_{D} = & -2\,\text{Re}\left[A_{S}^{}A_{P}\right] = \left|B_{1/2}\right|^{2} - \left|B_{-1/2}\right|^{2},\\
\beta_{D} = & -2\,\text{Im}\left[A_{S}^{}A_{P}\right] = 2\,\text{Im}\left[B_{1/2}B_{-1/2}^{}\right],\\
\gamma_{D} = & \left|A_{S}\right|^{2} - \left|A_{P}\right|^{2} = 2\,\text{Re}\left[B_{1/2}B_{-1/2}^{}\right],
\end{align}
where $\beta_{D} = \sqrt{1 - \alpha_{D}^{2}} \sin\phi_{D}$, and $\gamma_{D} = \sqrt{1 - \alpha_{D}^{2}} \cos\phi_{D}$. We give the explicit expressions for $a_{\mu\nu}$ in Appendix~\ref{E.spin transfer}. Contrary to common beliefs, $\alpha_{\Lambda}$ is not the most precise indicator of parity violation. Instead, $\gamma_{\Lambda}$ offers a  more direct measurement:  positive values  indicate parity violation dominance, while negative values suggest dominance of parity conservation. The key factor is $\phi_{\Lambda}$.  The Particle Data Group (PDG) reports $\phi_{\Lambda} = -6.5 \pm 3.5^{\circ}$~\cite{ParticleDataGroup:2022pth}, indicating a predominance of parity violation in $\Lambda$ decay.

The analysis of the decay for spin-3/2 particles, exemplified by the $\Omega^{-} \rightarrow \Lambda K^{-}$ process, parallels that for spin-1/2 particles. The spin density matrix of the $\Lambda$ particle is expressed as
\begin{align}
\rho_{1/2}^{\Lambda} = \sum_{\mu=0}^{15}\sum_{\nu=0}^{3}S_{\mu}b_{\mu\nu}\sigma_{\nu},\label{eq:Omega_decay}
\end{align}
where $S_{\mu}$ denotes the polarization of the parent particle $\Omega^{-}$, and $b_{\mu\nu}$, the polarization transfer matrix,  reflects the transfer of polarization from $\Omega^{-}$ to $\Lambda$. The matrix $b_{\mu\nu}$ is formulated as
\begin{align}
b_{\mu\nu} =& \frac{1}{2\pi}\sum_{\lambda,\lambda^{\prime}=-1/2}^{1/2}\sum_{\kappa,\kappa^{\prime}=-3/2}^{3/2} B_{\lambda}B_{\lambda^{\prime}}^{*}\left(\varSigma_{\mu}\right)^{\kappa,\kappa^{\prime}}\left(\sigma_{\nu}\right)^{\lambda^{\prime}\lambda}\mathcal{D}_{\kappa,\lambda}^{3/2*}\left(\Omega\right)\mathcal{D}_{\kappa^{\prime},\lambda^{\prime}}^{3/2}\left(\Omega\right),
\end{align}
where $B_{\lambda}$ is the helicity amplitude for the  $\Omega^{-} \rightarrow \Lambda K^{-}$ process with $\lambda$ being the helicity of the $\Lambda$. Using Eq.~\eqref{eq:canonical amplitudes}, these helicity amplitudes are described as:
\begin{align}
B_{-1/2}  = & \frac{\sqrt{2}}{2}\left(A_{P}+A_{D}\right),\label{eq:Bm1h_3/2}\\
B_{1/2}  = & \frac{\sqrt{2}}{2}\left(A_{P}-A_{D}\right).\label{eq:B1h__3/2}
\end{align}
Using Eq.~\eqref{eq:parity} for parity analysis, we find that that the $P$-wave term $(A_{P})$ corresponds to parity conservation, while the $D$-wave term $(A_{D})$ indicates parity-violating effects.  With the normalization condition $A_{P}^{2} + A_{D}^{2} = 1$, the amplitudes are parametrized as,
\begin{align}
\alpha_{D} = & -2\,\text{Re}\left[A_{P}^{*}A_{D}\right]=\left|B_{1/2}\right|^{2}-\left|B_{-1/2}\right|^{2}\label{eq:alpha_D_3/2},\\
\beta_{D} = & -2\,\text{Im}\left[A_{P}^{*}A_{D}\right]=2\,\text{Im}\left[B_{1/2}B_{-1/2}^{*}\right],\\
\gamma_{D} = & \left|A_{P}\right|^{2}-\left|A_{D}\right|^{2}=2\,\text{Re}\left[B_{1/2}B_{-1/2}^{*}\right],\label{eq:beta_D_3/2}
\end{align}
where $\beta_{D} = \sqrt{1-\alpha_{D}^{2}}\sin\phi_{D}$ and $\gamma_{D} = \sqrt{1-\alpha_{D}^{2}}\cos\phi_{D}$. The complete expressions for $b_{\mu\nu}$ are detailed in Appendix~\ref{E.spin transfer}. We use the parameter $\gamma_{\Omega}$ to assess the degree of parity violation, where a positive $\gamma_{\Omega}$ suggests a predominance of parity conservation, and a negative $\gamma_{\Omega}$ indicates  parity violation dominance. Commonly, decay parameters for $\Omega$ are expected to be $\beta_{\Omega} \approx 0$ and $\gamma_{\Omega} \approx \pm 1$~\cite{Luk:1988as,Kim:1992az}. Based on this, the STAR Collaboration reported $\gamma_{\Omega} \rightarrow 1$~\cite{STAR:2020xbm}. However, the BESIII Collaboration presented different results with, 
\begin{align}
&~  \beta_{\Omega}=-0.91_{-0.09}^{+0.21},\quad\gamma_{\Omega}=-0.41\pm0.46, \\
&~  \beta_{\Omega}=-0.85{}_{-0.15}^{+0.25},\quad\gamma_{\Omega}=-0.53\pm0.40,
\end{align}
for the two values of $\phi_{\Omega}$ in Solution I and Solutions II in Ref.~\cite{BESIII:2020lkm} together with $\alpha_{\Omega} = 0.0154$ ~\cite{ParticleDataGroup:2022pth} as inputs. These results significantly differ from traditional beliefs. The assumption that $\beta_{\Omega} \approx 0$, which implies limited time-reversal violation, is inconsistent with with the measurement of  BESIII Collaboration. Additionally, the sign of $\gamma_{\Omega}$ reported by BESIII Collaboration directly conflicts with the  result of STAR Collaboration. Given the notable experimental uncertainties, especially concerning $\phi_{\Omega}$,  more precise experimental investigations are needed to clarify these discrepancies.

Using the associated coordinate systems and angular definitions shown in Fig.~\ref{fig:Omega}, we have described decay processes using polarization transfer matrices $a_{\mu\nu}$ and $b_{\mu\nu}$,  While these matrices are computationally convenient, the lengthy polarization transfer expressions obscure the underlying physical mechanisms of decay processes. To provide a more intuitive understanding of decay processes, Lee and Yang proposed an alternate approach for spin-1/2 particle decays in Ref.~\cite{Lee:1957qs}. In the rest frame of the parent particle $\Lambda$, the polarization transfer in the $\Lambda \rightarrow p + \pi^{-}$ decay is expressed
\begin{align}
\vec{S}_{p} = \frac{\left(\alpha_{\Lambda}+\vec{S}_{\Lambda}\cdot\hat{p}_{p}\right)\hat{p}_{p}
+\beta_{\Lambda}\vec{S}_{\Lambda}\times\hat{p}_{p}+\gamma_{\Lambda}\hat{p}_{p}\times\left(\vec{S}_{\Lambda}\times\hat{p}_{p}\right)}
{1+\alpha_{\Lambda}\vec{S}_{\Lambda}\cdot\hat{p}_{p}},\label{eq:Lambda_decay2}
\end{align}
where $\vec{S}_{\Lambda}$ denotes the spin vector of $\Lambda$,  $\vec{p}_{p}$ and $\vec{S}_{p}$ represent the momentum and spin vector of the proton, respectively. The denominator indicates the cross-section for this process
\begin{align}
d\sigma^{\Lambda\rightarrow p\pi^{-}} \propto\left(1+\alpha_{\Lambda}\vec{S}_{\Lambda}\cdot\hat{p}_{p}\right).
\end{align}
This decay expression  clearly reveals the mechanisms involved. The parameter $\alpha_{\Lambda}$ reflects the impact of the parent polarization on both the cross-section and the polarization of decay products along their momentum direction.  $\beta_{\Lambda}$ details how the parent polarization influences the polarization of the decay products perpendicular to the plane formed by the parent spin vector and the momentum of the decay products. $\gamma_{\Lambda}$ signifies the effect of the parent polarization on the polarization of decay products within this plane, perpendicular to their momentum.

We extend this methodology to analyze the decay of spin-3/2 particles, using the $\Omega^- \rightarrow \Lambda \pi^-$ decay process as an example. According to the equivalence to  Eq.~\eqref{eq:Omega_decay}, the polarization of the $\Lambda$ particle is described as follows
\begin{align}
\vec{S}_{\Lambda}=&\frac{1}{D}\left\{ \left(\alpha_{\Omega}+\frac{2}{5}\vec{S}_{\Omega}\cdot\hat{p}_{\Lambda}-\alpha_{\Omega}T_{\Omega}^{pp}-2R_{\Omega}^{ppp}\right)\hat{p}_{\Lambda}\right.\nonumber\\
&\left.+\beta_{\Omega}\left(\frac{4}{5}\vec{S}_{\Omega}-2\vec{R}_{\Omega}^{pp}\right)\times\hat{p}_{\Lambda}+\gamma_{\Omega}\hat{p}_{\Lambda}\times\left[\left(\frac{4}{5}\vec{S}_{\Omega}-2\vec{R}_{\Omega}^{pp}\right)\times\hat{p}{\Lambda}\right]\right\},\label{eq:Omega_decay2}
\end{align}
where $D$ represents the cross-section term for the decay process
\begin{align}
d\sigma^{\Omega^{-}\rightarrow\Lambda K^{-}}\propto D=\left(1+\frac{2}{5}\alpha_{\Omega}\vec{S}_{\Omega}\cdot\hat{p}_{\Lambda}-T^{pp}_{\Omega}-2\alpha_{\Omega}R^{ppp}_{\Omega}\right),
\end{align}
where $S^{i}_{\Omega}$, $T^{ij}_{\Omega}$, and $R^{ijk}_{\Omega}$ denote the spin vector, rank-2 spin tensor, and rank-3 spin tensor of the $\Omega^-$ particle, detailed in Appendix~\ref{B.Probabilistic}. The following shorthand notations are used in the expressions,
\begin{align}
T^{pp}_{\Omega}= & T^{ij}_{\Omega}\hat{p}_{\Lambda}^{i}\hat{p}_{\Lambda}^{j},\\
R^{ppp}_{\Omega}= & R^{ijk}_{\Omega}\hat{p}_{\Lambda}^{i}\hat{p}_{\Lambda}^{j}\hat{p}_{\Lambda}^{k},\\
\left(R^{pp}_{\Omega}\right)^{i}= & R^{ijk}_{\Omega}\hat{p}_{\Lambda}^{j}\hat{p}_{\Lambda}^{k}.
\end{align}
Similar to the decay of spin-1/2 particles, the decay parameters $\alpha_{\Omega}$, $\beta_{\Omega}$, and $\gamma_{\Omega}$ for spin-3/2 particles also carry related physical interpretations. Our approach shows some differences from the one in Ref.~\cite{Kim:1992az}, mainly due to a distinct normalization method for polarization components. We detail the domains of the polarization components based on our normalization scheme in Eq.~\eqref{eq:spin_range}. A direct difference is seen when averaging over the angular distribution of $\Lambda$, where the polarization transfer is simplified to
\begin{align}
\vec{P}_{\Lambda} = C_{\Omega\Lambda}\left(\frac{2}{3}\vec{S}_{\Omega}\right) = \frac{1}{5}\left(1 + 4\gamma_{\Omega}\right)\left(\frac{2}{3}\vec{S}_{\Omega}\right).
\end{align}
Our formula includes a factor of $2/3$, which is absent in Refs.~\cite{Kim:1992az,STAR:2020xbm}, which arises from normalizing the spin vector $S^{i}$ in the range of $[-3/2, 3/2]$.

We has explored two different methods for describing decay processes. The helicity formalism is particularly useful for depicting the coordinate system and angular dependencies of the decay products. On the other hand, the Lee-Yang method offers more explicit insights into the fundamental physical principles of these processes. Combining the advantages of these two approaches, we develop the third method to describe decay. For the decay of spin-1/2 particles, such as in the $\Lambda \rightarrow p + \pi^{-}$ process, we establish the coordinate system of the parent particle as $x_{\Lambda}$-$y_{\Lambda}$-$z_{\Lambda}$, illustrated in Fig.~\ref{fig:Omega}. The polarization projection axes for the proton are represented as,
\begin{align}
\hat{x}_{p} & =\left\{ \cos\theta_{p}\cos\phi_{p},\cos\theta_{p}\sin\phi_{p},-\sin\theta_{p}\right\} ,\label{eq:x_p}\\
\hat{y}_{p} & =\left\{ -\sin\phi_{p},\cos\phi_{p},0\right\} ,\\
\hat{z}_{p} & =\left\{ \sin\theta_{p}\cos\phi_{p},\sin\theta_{p}\sin\phi_{p},\cos\theta_{p}\right\}.\label{eq:z_p}
\end{align}
By projecting Eq.~\eqref{eq:Lambda_decay2} onto these axes, we determine the cross-section and polarization components of the proton for each axis,
\begin{align}
d\sigma^{\Lambda \rightarrow p\pi^{-}} & \propto D = (1 + \alpha_{\Lambda}S_{\Lambda}^{i}\hat{z}_{p}^{i}), \\
P_{x}^{p} & = \frac{1}{D} S_{\Lambda}^{i}(\beta_{\Lambda}\hat{y}_{p}^{i} + \gamma_{\Lambda}\hat{x}_{p}^{i}), \\
P_{y}^{p} & = \frac{1}{D} S_{\Lambda}^{i}(\gamma_{\Lambda}\hat{y}_{p}^{i} - \beta_{\Lambda}\hat{x}_{p}^{i}), \\
P_{z}^{p} & = \frac{1}{D} (\alpha_{\Lambda} + S_{\Lambda}^{i}\hat{z}_{p}^{i}).
\end{align}
This representation not only includes details about the coordinate axes and angles but also offers a clearer understanding of the physical mechanisms involved. Furthermore, by decomposing polarization into individual axes, this approach simplifies the study of polarization in various directions.

We apply this method to express the decay of spin-3/2 particles. Using a similar approach to define the coordinate system for the parent particle $\Omega$ as $\hat{x}_{\Omega}$-$\hat{y}_{\Omega}$-$\hat{z}_{\Omega}$, and the polarization projection axes for the daughter particle $\Lambda$ as $\hat{x}_{\Lambda}$-$\hat{y}_{\Lambda}$-$\hat{z}_{\Lambda}$, we project Eq.~\eqref{eq:Omega_decay2} onto these axes, and obtain,
\begin{align}
d\sigma^{\Omega^{-}\rightarrow\Lambda K^{-}} \propto & D = \left(1+\frac{2}{5}\alpha_{\Omega}S_{\Omega}^{i}\hat{z}_{\Lambda}^{i}-T_{\Omega}^{ij}\hat{z}_{\Lambda}^{i}\hat{z}_{\Lambda}^{j}-2\alpha_{\Omega}R_{\Omega}^{ijk}\hat{z}_{\Lambda}^{i}\hat{z}_{\Lambda}^{j}\hat{z}_{\Lambda}^{k}\right),\\
P_{x}^{\Lambda} = & \frac{1}{D}\left[\frac{4}{5}S_{\Omega}^{i}\left(\beta_{\Omega}\hat{y}_{\Lambda}^{i}+\gamma_{\Omega}\hat{x}_{\Lambda}^{i}\right)-2R_{\Omega}^{ijk}\hat{z}_{\Lambda}^{i}\hat{z}_{\Lambda}^{j}\left(\beta_{\Omega}\hat{y}_{\Lambda}^{k}+\gamma_{\Omega}\hat{x}_{\Lambda}^{k}\right)\right],\\
P_{y}^{\Lambda} = & \frac{1}{D}\left[\frac{4}{5}S_{\Omega}^{i}\left(\gamma_{\Omega}\hat{y}_{\Lambda}^{i}-\beta_{\Omega}\hat{x}_{\Lambda}^{i}\right)-2R_{\Omega}^{ijk}\hat{z}_{\Lambda}^{i}\hat{z}_{\Lambda}^{j}\left(\gamma_{\Omega}\hat{y}_{\Lambda}^{k}-\beta_{\Omega}\hat{x}_{\Lambda}^{k}\right)\right],\\
P_{z}^{\Lambda} = & \frac{1}{D}\left[\alpha_{\Omega}+\frac{2}{5}S_{\Omega}^{i}\hat{z}_{\Lambda}^{i}-\alpha_{\Omega}T_{\Omega}^{ij}\hat{z}_{\Lambda}^{i}\hat{z}_{\Lambda}^{j}-2R_{\Omega}^{ijk}\hat{z}_{\Lambda}^{i}\hat{z}_{\Lambda}^{j}\hat{z}_{\Lambda}^{k}\right].
\end{align}

Using these decay expressions,  we establish the joint angular distributions for final-state particles in single-tag $\Omega^{-}$ and double-tag $\Omega^{-}\bar{\Omega}^{+}$ decays. Due to the equivalence of the three polarization transfer expressions, we focus on the first method for computational convenient.

For single-tag $\Omega^{-}$ decays, the joint angular distribution is formulated as
\begin{align}
\mathcal{W}\left(\vec{\omega},\vec{\zeta}\right) = \sum_{\mu=0}^{15}\sum_{\nu=0}^{3}S_{\mu}b_{\mu\nu}^{\Omega}a_{\nu0}^{\Lambda},\label{eq:single_cross}
\end{align}
where $\vec{\omega}=\left\{\alpha_{\psi},\alpha_{1},\alpha_{2},\phi_{1},\phi_{3},\phi_{4},\alpha_{\Omega},\phi_{\Omega},\alpha_{\Lambda}\right\} $  denotes the decay parameters, and $\vec{\xi}=\left\{ \theta_{\Omega},\theta_{\Lambda},\phi_{\Lambda},\theta_{p},\phi_{p}\right\} $  indicates the angles involved in both the production and multi-stage decay processes. Considering the challenges in directly measuring the polarization of the proton, our analysis includes a summation over this polarization..

For double-tag $\Omega^{-}\bar{\Omega}^{+}$  decays, we express the joint angular distribution as follows
\begin{align}
\mathcal{W}\left(\vec{\omega},\vec{\zeta}\right) & =\sum_{\mu=0}^{15}\sum_{\nu=0}^{15}\sum_{\mu^{\prime}=0}^{3}\sum_{\nu^{\prime}=0}^{3}S_{\mu\nu}b_{\mu\mu^{\prime}}^{\Omega}b_{\nu\nu^{\prime}}^{\bar{\Omega}}a_{\mu^{\prime}0}^{\Lambda}a_{\nu^{\prime}0}^{\bar{\Lambda}},\label{eq:double_cross}
\end{align}
where $\vec{\omega}=\left\{\alpha_{\psi},\alpha_{1},\alpha_{2},\phi_{1},\phi_{3},\phi_{4},\alpha_{\Omega},\alpha_{\bar{\Omega}},\phi_{\Omega},\phi_{\bar{\Omega}},\alpha_{\Lambda},\alpha_{\bar{\Lambda}}\right\}$ signifies decay parameters, and $\vec{\xi}=\left\{\theta_{\Omega},\theta_{\Lambda},\phi_{\Lambda},\theta_{\bar{\Lambda}},\phi_{\bar{\Lambda}},\theta_{p},\phi_{p},\theta_{\bar{p}},\phi_{\bar{p}}\right\}$ reflects the associated angles. Similar to the single-tag case, we sum over the proton polarization.

\section{Further Discussion: Sensitivity of $\phi_{\Omega}$ measurement} \label{s.sensitivity}

In this section, we compare the sensitivity of  the decay parameter $\phi_{\Omega}$ in single-tag $\Omega^{-}$ decays and double-tag $\Omega^{-}\bar{\Omega}^{+}$ decays measurements. Following the methods outlined in Refs.~\cite{Han:2019axh,Hong:2022prk}, we use the maximum likelihood method to examine the sensitivity of $\phi_{\Omega}$  with respect to  the number of observed events, $N$.

For a specific process, we define the normalized joint angular distribution as
\begin{align}
\widetilde{\mathcal{W}}\left(\vec{\omega},\vec{\zeta}\right)= & \frac{\mathcal{W}(\vec{\omega},\vec{\zeta})}{\int\mathcal{W}\left(\vec{\omega},\vec{\zeta}\right)d\vec{\zeta}},\label{eq:normalized_cross_section}
\end{align}

For a set of specific data, the likelihood function can be defined as
\begin{align}
\mathcal{L}= & \prod_{i=1}^{N}\widetilde{\mathcal{W}}\left(\vec{\omega},\vec{\zeta}\right),\label{eq:likelihood}
\end{align}
where $N$ is the number of the observed events. In the maximum likelihood method, the statistical sensitivity of the measured parameter is determined by the relative uncertainty,
\begin{align}
\delta\left(\phi_{\Omega}\right)= & \frac{\sqrt{V\left(\phi_{\Omega}\right)}}{\left|\phi_{\Omega}\right|},\label{eq:sensitivity}
\end{align}
where $V\left(\phi_{\Omega}\right)$ is the variance of the parameter, given by
\begin{align}
V^{-1}\left(\phi_{\Omega}\right)= & N\int\frac{1}{\widetilde{\mathcal{W}}}\left[\frac{\partial\widetilde{\mathcal{W}}}{\partial\phi_{\Omega}}\right]^{2}d\vec{\zeta}.\label{eq:variance}
\end{align}
To determine how the sensitivity of the parameter $\phi_{\Omega}$ depends on the number of observed events $N$ in both single-tag and double-tag processes, we insert Eqs.~\eqref{eq:single_cross} and ~\eqref{eq:double_cross} into Eqs.~\eqref{eq:normalized_cross_section}-\eqref{eq:variance}. We set the polarization-related parameters as follows~\cite{BESIII:2020lkm,ParticleDataGroup:2022pth},
\begin{align}
\begin{centering}
\begin{array}{ccc}
\alpha_{\psi}= 0.237, & \alpha_{1}=-0.371, & \alpha_{2}=1.090,\\
\phi_{1}= 4.37,& \phi_{3}=2.60,& \phi_{4}=4.02,\\
\alpha_{\Lambda/\bar{\Lambda}}=\pm0.753,& \alpha_{\Omega/\bar{\Omega}}=\pm0.0154,& \phi_{\Omega/\bar{\Omega}}=\pm4.22.
\end{array}
\end{centering}
\label{eq:parameter_value}
\end{align}
For this preliminary assessment, we consider only the central values.

\begin{figure}
\begin{centering}
\includegraphics[width=5.5cm,height=5cm]{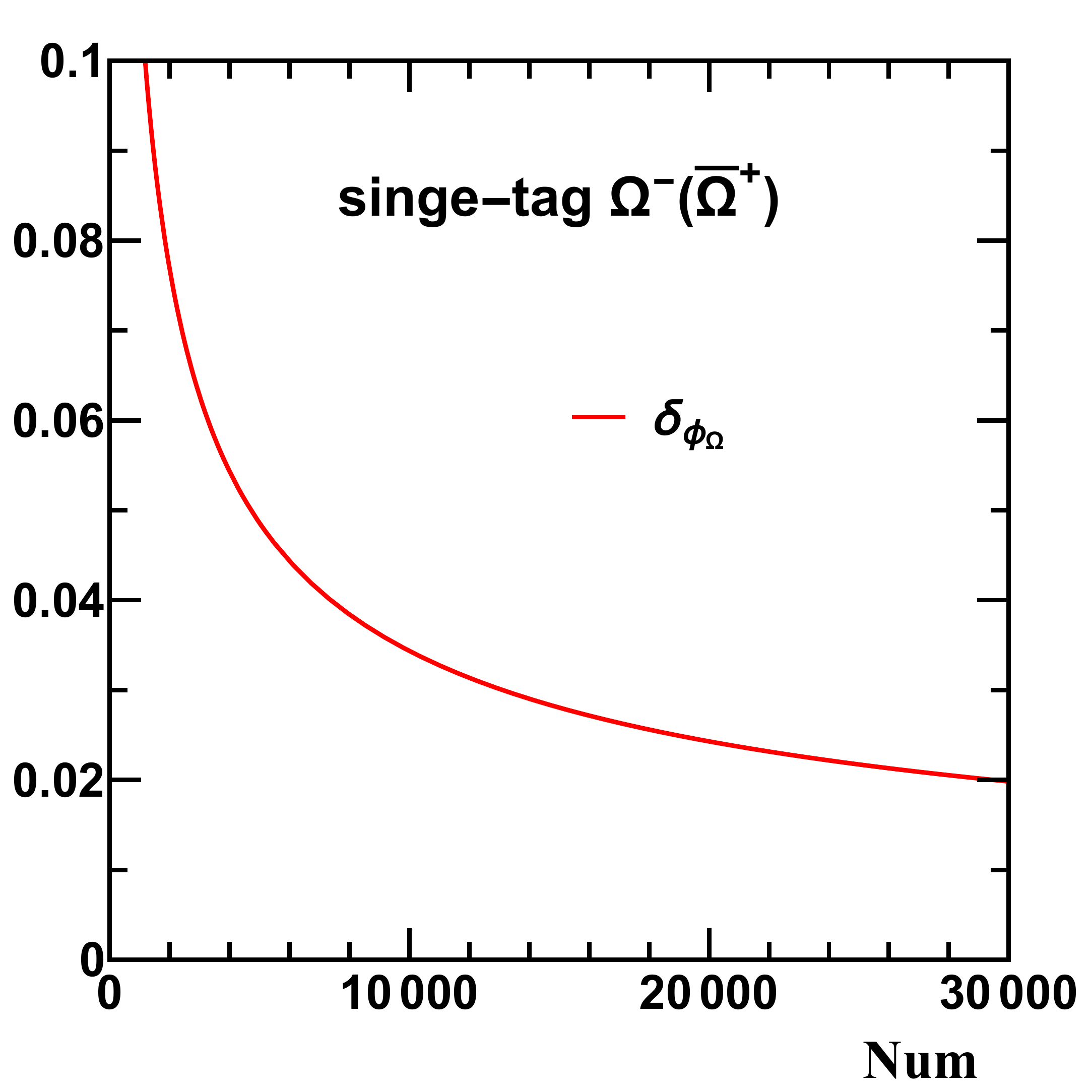}\includegraphics[width=5.5cm,height=5cm]{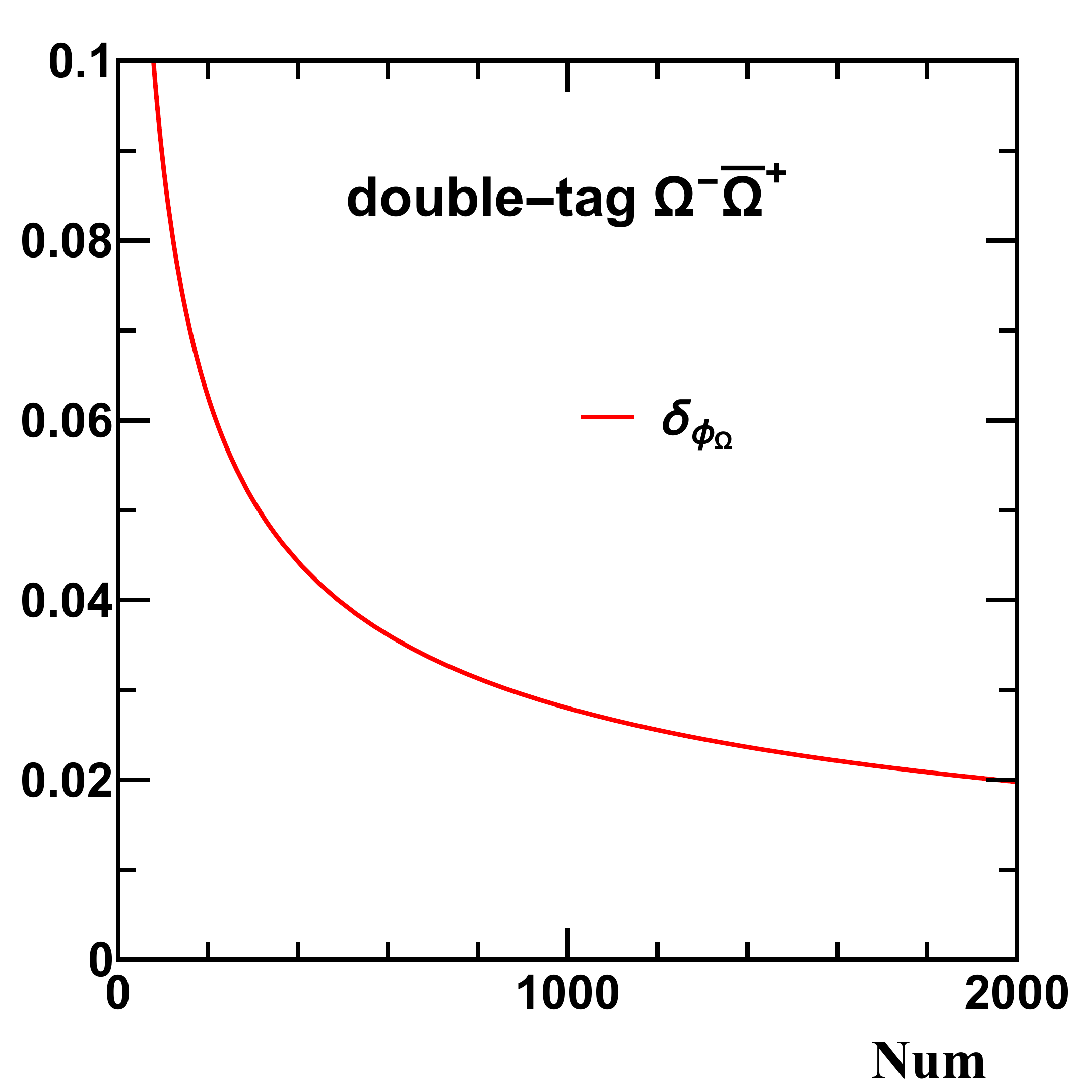}
\par\end{centering}
\caption{\label{fig:sensitivity}Statistical sensitivity of $\phi_\Omega$ estimated via maximum likelihood estimation. The sensitivity under single-tag (left) and double-tag (right) conditions is plotted as a function of the number of events $N$. For equivalent statistical sensitivity, double-tag events require only 6-7\% of the number of single-tagged events.}
\end{figure}

We present the statistical sensitivity of the parameter $\phi_{\Omega}$ in Fig.~\ref{fig:sensitivity}. This prediction does not account for parameters uncertainties, background effects, detection efficiency, angular acceptance, or various systematic uncertainties. Consequently, the sensitivity depicted in Fig.~\ref{fig:sensitivity} is likely to be an overestimate compared to actual experimental measurement uncertainties. Our focus here is on providing rough estimates and comparative analyses.

Fig.~\ref{fig:sensitivity} indicates that for a similar level of statistical sensitivity, double-tag measurements require only about 6-7\% of the number of events needed in single-tag measurements. This implies that double-tag becomes statistically more efficient when the detection efficiency for single-side $\Omega$ decay exceeds 15\%. Current reports suggest detection efficiencies for single-side $\Omega$ decays are around 17-19\%~\cite{BESIII:2020lkm}. Moreover, single-tag $\Omega^{-}$ measurements typically include a background of approximately 10\%~\cite{BESIII:2020lkm}, which does not decrease with larger sample sizes.  By contrast, double-tag measurements are likely to have a much lower background, estimated to be under 0.5\%. This reduced background is mainly due to the proximity of $\psi(3686)$ to the $\Omega^{-}\bar{\Omega}^{+}$ production threshold.

Overall, double-tag measurement offers superior performance in all aspects. Given the current detection efficiencies and the recent accumulation of events at $\psi(3686)$ ~\cite{BESIII:2023olq}, we anticipate gathering around 1600-1800 double-tag events. As shown in Fig.~\ref{fig:sensitivity}, this amount of data could lead to a sensitivity of approximately 2\% for $\phi_{\Omega}$. Such improved sensitivity is expected to clarify the current discrepancies in the measurements of $\phi_\Omega$ among different experiments.

\section{Summary} \label{s.summary}

In this paper, we perform a polarization-related physical analysis in the $e^{+}e^{-}\rightarrow\Omega^{-}\bar{\Omega}^{+}$ process, covering the following aspects: the production density matrix for $\Omega^{-}\bar{\Omega}^{+}$ pairs, single-tag $\Omega^{-}$ polarization expansion coefficients, double-tag $\Omega^{-}\bar{\Omega}^{+}$ polarization correlation coefficients, and the decay chains of $\Omega^{-}\bar{\Omega}^{+}$.

Using the helicity formalism, we present the production density matrix for $\Omega^{-}\bar{\Omega}^{+}$ pairs . This matrix is defined by four complex amplitudes: $H_{1}$, $H_{2}$, $H_{3}$, and $H_{4}$. It adheres to the fundamental principles of parity conservation and charge conjugation invariance. These amplitudes provide insights into the polarization properties and form factors of $\Omega^-$ particles. Investigating the form factors in the timelike region contributes to the understanding of the internal structure of $\Omega^{-}$ and serves as a vital reference for theoretical approaches such as lattice QCD and other nucleon structure models.

We review the representation of spins for particles with 1/2 and 3/2 spins, discussing both the general formalism for spin density matrices and their representation within the helicity formalism. The general formalism provides a comprehensive interpretation of spin components, while the helicity formalism offers a more concise representation but lacks intuitive physical insights. To bridge these methodologies, we introduced a new basis matrix within the helicity formalism, which correlates spin components defined in the helicity formalism with those in the general spin density matrix. This approach facilitates a deeper understanding of the underlying spin mechanics in physical processes.

We introduce a novel parametrization for helicity amplitudes  to investigate the polarization properties in both single-tag $\Omega^{-}$ and double-tag $\Omega^{-}\bar{\Omega}^{+}$  cases. For the single-tag $\Omega^{-}$,  we identify six non-zero polarization components, discuss their physical interpretations, and explain why only these components are feasible. By establishing parameters range values, we give the  domains of these polarization components. Notably, we identify specific sets of solutions that can render the $\Omega^{-}$ particle unpolarized. 

 In the case of double-tag $\Omega^{-}\bar{\Omega}^{+}$, 116 out of the 256 potential polarization correlation coefficients are non-zero.  We classify these coefficients into exchange symmetric and exchange antisymmetric terms  in the exchange of $\mu$ and $\nu$. The presence of these symmetric and antisymmetric terms is a result of CP conservation principles. Furthermore, we identifiy a specific set of solutions that minimizes polarization correlation in $\Omega^{-}\bar{\Omega}^{+}$ pairs. Double-tag measurements offer the advantage of eliminating non-physical solutions that may exist in single-tag measurements.

Particle polarization analysis often relies on studying their decay processes. We present three equivalent approaches to represent the decay of particles with spin 1/2 and 3/2, providing valuable insights into these decay mechanisms from different perspectives. Notably, current experimental data reveal inconsistencies in the decay parameters of the $\Omega$  particle, which present substantial challenges in the study of processes involving $\Omega$. This highlights the  necessity for more precise experimental observations to resolve these discrepancies.

The inconsistencies in the understanding of the $\Omega$ particle decay parameters center around $\phi_{\Omega}$. By employing maximum likelihood estimation, we  assess the statistical sensitivity of $\phi_{\Omega}$ with respect to  the number of observed events ($N$) in both single-tag and double-tag measurements. Taking into account the reported $\Omega$ particle reconstruction efficiency by the BESIII collaboration, our findings indicate that the double-tag measurement offers statistical advantages. Additionally, this approach effectively reduces background noise. Based on the data collected by BESIII at $\psi(3686)$, our predictions suggest that double-tag measurements can constrain the statistical uncertainty of $\phi_{\Omega}$ to approximately 2\%, thereby resolving the existing discrepancies related to this decay phase.

In conclusion, our study offers fresh insights into the polarization phenomena during the production of $\Omega^{-}\bar{\Omega}^{+}$ pairs in $e^{+}e^{-}$ annihilation processes. The analytical framework we have developed is not limited to $\Omega^{-}\bar{\Omega}^{+}$ pairs but is also applicable to the polarization analysis of other spin-3/2 particle pairs produced in electron-positron annihilation.  The decay formalism established for spin-3/2 particles holds universal applicability in a wide range of research areas involving such particles, including the investigation of spin-3/2 fragmentation functions, global polarization, and related topics.

\acknowledgments{
The authors would like to thank Tianbo Liu, Zuo-tang Liang, Ronggang Ping, Weihua Yang, and Ruoyu Zhang for useful discussions. This work is Supported by the National Natural Science Foundation of China (Approvals No. 12247121, and No. 12305085) and the Shandong Province Natural Science Foundation under Grant No. ZR2020MA098.}

\appendix

\section{form factors in time-like region}\label{A.form_factors}

The relationship between the form factors and the transition amplitudes is given by~\cite{Korner:1976hv,Perotti:2018wxm}:
\begin{align}
G_{E} = & \frac{1}{4m}\left[H_{1}+H_{4}\right],\label{eq:GE}\\
G_{M} = & \frac{3}{5\sqrt{2}}\frac{1}{\sqrt{q^{2}}}\left[\sqrt{3}H_{3}+H_{2}\right],\label{eq:GM}\\
G_{Q} = & \frac{3}{2\tau}\frac{1}{4m}\left[H_{4}-H_{1}\right],\label{eq:GQ}\\
G_{O} = & \frac{1}{4\tau}\frac{1}{\sqrt{q^{2}}}\left[H_{3}-\frac{\sqrt{3}}{2}H_{2}\right],\label{eq:GO}
\end{align}
where $\tau=q^{2}/4m^{2}$, $m$ is the mass of the $\Omega^{-}$, and $q$ is the momentum of the virtual photon.

\begin{table}[ht]
\caption{\label{tab:ratios_form_factor}The ratios of form factors for $\Omega$ particle.  These values are derived using Eqs.~\eqref{eq:GE}-\eqref{eq:GO} , based on BESIII Collaboration measurements~\cite{BESIII:2020lkm}.}
\renewcommand\arraystretch{2}
\[
\begin{tabular}{c|c|c}
\hline\hline  ~~Ratio~~ & ~~~~~Solution I~~~~~ & ~~~~~Solution II~~~~~\\
\hline  $\frac{\left|G_{M}\right|}{\left|G_{E}\right|}$ & $0.632\pm0.151$ & $0.698\pm0.201$\\
\hline  $\frac{\left|G_{Q}\right|}{\left|G_{E}\right|}$ & $0.386\pm0.336$ & $0.737\pm0.613$\\
\hline  $\frac{\left|G_{O}\right|}{\left|G_{E}\right|}$ & $0.261\pm0.082$ & $0.321\pm0.159$\\
\hline\hline \end{tabular}
\]
\end{table}

Using the data sets in Ref.~\cite{BESIII:2020lkm}, we present the ratios of form factors in Table~\ref{tab:ratios_form_factor}. Although these measurements have limited precision, they continue to be valuable references for studies in lattice QCD and quark models, particularly concerning the form factors of the $\Omega^{-}$ baryon.

\section{Spin compenents for spin 3/2 and their physical interpretation}\label{B.Probabilistic}

The polarization of spin-3/2 particles can be described using 15 independent spin components.  These components are found in the spin vector $S^{i}$, the rank-two spin tensor $T^{ij}$, and the rank-three spin tensor $R^{ijk}$. Together, they constitute the spin density matrix that represents a spin-3/2 particle. The physical interpretations of these components are expressed as combinations of probabilities to find specific polarization states in the system. We provide the explicit expressions of $S^{i}$ , $T^{ij}$ , and $R^{ijk}$ in the particle rest frame and the corresponding physical interpretations for all 15 polarization components.  It is worth noting that our decomposition of the physical interpretation for $S_{LTT}^{xy}$ differs from the one presented in Refs.~\cite{Zhao:2022lbw,Zhang:2023wmd}, allowing for a  refined analysis of parity or CP symmetries.

For spin vector $S^{i}$, which consists of three polarization components, the expression is given by
\begin{align}
S^{i}=  \left(S_{T}^{x},S_{T}^{y},S_{L}\right).\label{eq:spin_vector_3/2}
\end{align}
These polarization components are defined using $\left\langle \Sigma^{a}\right\rangle=\text{Tr}\left[\Sigma^{a}\rho\right]$ with $\Sigma^{a}$ defined in Eqs.~\eqref{eq:spin_vector}-\eqref{eq:spin_tensor2}. Then, these polarization components are represented as,
\begin{align}
S_{L}= \langle\Sigma^{z}\rangle,\quad S_{T}^{x}=\langle\Sigma^{x}\rangle,\quad S_{T}^{y}=\langle\Sigma^{y}\rangle.\label{eq:component1}
\end{align}
We will provide the physical interpretations of these components by exploring this definition.

For the rank-two spin tensor $T^{ij}$, which consists of five polarization components, the expression is given by
\begin{align}
T^{ij} & =\frac{1}{2}\begin{pmatrix}-S_{LL}+S_{TT}^{xx} & S_{TT}^{xy} & S_{LT}^{x}\\
S_{TT}^{xy} & -S_{LL}-S_{TT}^{xx} & S_{LT}^{y}\\
S_{LT}^{x} & S_{LT}^{y} & 2S_{LL}
\end{pmatrix},\label{eq:spin_tensor_3/2}
\end{align}
These polarization components are represented as,
\begin{align}
\begin{aligned}S_{LL} & =\langle\Sigma^{zz}\rangle,\quad S_{LT}^{x}=2\langle\Sigma^{xz}\rangle,\quad S_{LT}^{y}=2\langle\Sigma^{yz}\rangle,\\
S_{TT}^{xy} & =2\langle\Sigma^{xy}\rangle,\quad S_{TT}^{xx}=\langle\Sigma^{xx}-\Sigma^{yy}\rangle.
\end{aligned}
\label{eq:components2}
\end{align}

For the rank-three spin tensor $R^{ijk}$, which consists of seven polarization components, the expression is given by
\begin{align}
R^{ijk}= & \frac{1}{4}\left[\begin{array}{c}
\left(\begin{array}{ccc}
-3S_{LLT}^{x}+S_{TTT}^{xxx} & -S_{LLT}^{y}+S_{TTT}^{yxx} & -2S_{LLL}+S_{LTT}^{xx}\\
-S_{LLT}^{y}+S_{TTT}^{yxx} & -S_{LLT}^{x}-S_{TTT}^{xxx} & S_{LTT}^{xy}\\
-2S_{LLL}+S_{LTT}^{xx} & S_{LTT}^{xy} & 4S_{LLT}^{x}
\end{array}\right)\\
\left(\begin{array}{ccc}
-S_{LLT}^{y}+S_{TTT}^{yxx} & -S_{LLT}^{x}-S_{TTT}^{xxx} & S_{LTT}^{xy}\\
-S_{LLT}^{x}-S_{TTT}^{xxx} & -3S_{LLT}^{y}-S_{TTT}^{yxx} & -2S_{LLL}-S_{LTT}^{xx}\\
S_{LTT}^{xy} & -2S_{LLL}-S_{LTT}^{xx} & 4S_{LLT}^{y}
\end{array}\right)\\
\left(\begin{array}{ccc}
-2S_{LLL}+S_{LTT}^{xx} & S_{LTT}^{xy} & 4S_{LLT}^{x}\\
S_{LTT}^{xy} & -2S_{LLL}-S_{LTT}^{xx} & 4S_{LLT}^{y}\\
4S_{LLT}^{x} & 4S_{LLT}^{y} & 4S_{LLL}
\end{array}\right)
\end{array}\right].\label{eq:spin_tensor2_3/2}
\end{align}
These polarization components are represented as,
\begin{align}
\begin{aligned}
S_{LLL} =&\langle\Sigma^{zzz}\rangle,\quad S_{LLT}^{x}=\langle\Sigma^{xzz}\rangle,\quad S_{LLT}^{y}=\langle\Sigma^{yzz}\rangle,\\
\quad S_{LTT}^{xy}=&4\langle\Sigma^{xyz}\rangle,S_{LTT}^{xx} =2\langle\Sigma^{xxz}-\Sigma^{yyz}\rangle,\\
\quad S_{TTT}^{xxx}=&\langle\Sigma^{xxx}-3\Sigma^{xyy}\rangle,\quad S_{TTT}^{yxx}=\langle3\Sigma^{yxx}-\Sigma^{yyy}\rangle.
\end{aligned}
\label{eq:component3}
\end{align}

In order to analyze the polarization states, we introduce the spin projection operators along a specific direction $(\theta,\phi)$ as follows
\begin{align}
\Sigma^{i}\hat{n}_{i}=\Sigma^{x}\sin\theta\cos\phi+\Sigma^{y}\sin\theta\sin\phi+\Sigma^{z}\cos\theta.
\end{align}
where $\theta$ represents the polar angle of this direction, and $\phi$ represents the azimuthal angle. We define the eigenstates along this particular direction as $\left|m_{\left(\theta,\phi\right)}\right\rangle $, where $m$ denotes the corresponding eigenvalue. Therefore, the probability of finding this polarization state in the system is given by
\begin{align}
P\left(m_{(\theta,\phi)}\right)=\text{Tr}\left[\rho|m_{(\theta,\phi)}\rangle\langle m_{(\theta,\phi)}|\right].\label{eq:probability}
\end{align}

To facilitate our analysis, we introduce the following notations,
\begin{align}
\begin{aligned}
|m\rangle_{x+y}= & \left|m_{(\frac{\pi}{2},\frac{\pi}{4})}\right\rangle ,\quad|m\rangle_{x+z}=\left|m_{\left(\frac{\pi}{4},0\right)}\right\rangle,\\
|m\rangle_{y+z}=&\left|m_{\left(\frac{\pi}{4},\frac{\pi}{2}\right)}\right\rangle,|m\rangle_{x-y}= \left|m_{(\frac{\pi}{2},-\frac{\pi}{4})}\right\rangle,\\
|m\rangle_{x-z}=&\left|m_{\left(-\frac{\pi}{4},0\right)}\right\rangle ,\quad|m\rangle_{y-z}=\left|m_{\left(-\frac{\pi}{4},\frac{\pi}{2}\right)}\right\rangle ,\\
|m\rangle_{x+y+z}= & \left|m_{\left(\theta_{xyz},\frac{\pi}{4}\right)}\right\rangle ,\quad|m\rangle_{x-y+z}=\left|m_{\left(\theta_{xyz},-\frac{\pi}{4}\right)}\right\rangle ,\\
|m\rangle_{x+y-z}= & \left|m_{\left(\pi-\theta_{xyz},\frac{\pi}{4}\right)}\right\rangle ,\quad|m\rangle_{x-y-z}=\left|m_{\left(\pi-\theta_{xyz},-\frac{\pi}{4}\right)}\right\rangle,
\end{aligned}
\label{eq:short_notation_xyz}
\end{align}
where $\theta_{xyz}$=arctan($\sqrt{2}$). These notations will be useful in understanding the symmetry of the reaction process.

Accoding to Eq.~\eqref{eq:spin_vector_3/2} and Eq.~\eqref{eq:component1}, the physical interpretations of the three spin vector components are given by,
\begin{align}
S_{L}= & \frac{3}{2}\left[P_{z}\left(\frac{3}{2}\right)-P_{z}\left(-\frac{3}{2}\right)\right]+\frac{1}{2}\left[P_{z}\left(\frac{1}{2}\right)-P_{z}\left(-\frac{1}{2}\right)\right],\label{eq:SL_interpretation}\\
S_{T}^{x}= & \frac{3}{2}\left[P_{x}\left(\frac{3}{2}\right)-P_{x}\left(-\frac{3}{2}\right)\right]+\frac{1}{2}\left[P_{x}\left(\frac{1}{2}\right)-P_{x}\left(-\frac{3}{2}\right)\right],\\
S_{T}^{y}= & \frac{3}{2}\left[P_{y}\left((\frac{3}{2}\right)-P_{y}\left(-\frac{3}{2}\right)\right]+\frac{1}{2}\left[P_{y}\left(\frac{1}{2}\right)-P_{y}\left(-\frac{1}{2}\right)\right].
\end{align}

Accoding to Eq.~\eqref{eq:spin_tensor_3/2} and Eq.~\eqref{eq:components2}, the physical interpretations of the five rank-two spin tensor components are given by,
\begin{align}
S_{LL}= & \left[P_{z}\left(\frac{3}{2}\right)+P_{z}\left(-\frac{3}{2}\right)\right]-\left[P_{z}\left(\frac{1}{2}\right)+P_{z}\left(-\frac{1}{2}\right)\right],\\
S_{LT}^{x}= & 2\left[P_{x+z}\left(\frac{3}{2}\right)+P_{x+z}\left(-\frac{3}{2}\right)\right]-2\left[P_{x-z}\left(\frac{3}{2}\right)+P_{x-z}\left(-\frac{3}{2}\right)\right],\\
S_{LT}^{y}= & 2\left[P_{y+z}\left(\frac{3}{2}\right)+P_{y+z}\left(-\frac{3}{2}\right)\right]-2\left[P_{y-z}\left(\frac{3}{2}\right)+P_{y-z}\left(-\frac{3}{2}\right)\right],\\
S_{TT}^{xx}= & 2\left[P_{x}\left((\frac{3}{2}\right))+P_{x}\left((-\frac{3}{2}\right))\right]-2\left[P_{y}\left((\frac{3}{2}\right))+P_{y}\left((-\frac{3}{2}\right))\right],\\
S_{TT}^{xy}= & 2\left[P_{x+y}\left((\frac{3}{2}\right))+P_{x+y}\left((-\frac{3}{2}\right))\right]-2\left[P_{x-y}\left((\frac{3}{2}\right))+P_{x-y}\left((-\frac{3}{2}\right))\right].
\end{align}

Accoding to Eq.~\eqref{eq:spin_tensor2_3/2} and Eq.~\eqref{eq:component3}, the physical interpretations of the seven rank-three spin tensor components are given by,
\begin{align}
S_{LLL}= & \frac{3}{10}\left[P_{z}\left(\frac{3}{2}\right)-P_{z}\left(-\frac{3}{2}\right)\right]-\frac{9}{10}\left[P_{z}\left(\frac{1}{2}\right)-P_{z}\left(-\frac{1}{2}\right)\right],\\
S_{LLT}^{x}= & -\frac{1}{60}\left\{ 129\left[P_{x}\left(\frac{3}{2}\right)-P_{x}\left(-\frac{3}{2}\right)\right]+23\left[P_{x}\left(\frac{1}{2}\right)-P_{x}\left(-\frac{1}{2}\right)\right]\right\} \nonumber \\
 & +\frac{\sqrt{2}}{24}\left\{ 27\left[P_{x+z}\left(\frac{3}{2}\right)-P_{x+z}\left(-\frac{3}{2}\right)\right]+\left[P_{x+z}\left(\frac{1}{2}\right)-P_{x+z}\left(-\frac{1}{2}\right)\right]\right\} \nonumber \\
 & +\frac{\sqrt{2}}{24}\left\{ 27\left[P_{x-z}\left(\frac{3}{2}\right)-P_{x-z}\left(-\frac{3}{2}\right)\right]+\left[P_{x-z}\left(\frac{1}{2}\right)-P_{x-z}\left(-\frac{1}{2}\right)\right]\right\} ,\\
S_{LLT}^{y}= & -\frac{1}{60}\left\{ 129\left[P_{y}\left(\frac{3}{2}\right)-P_{y}\left(-\frac{3}{2}\right)\right]+23\left[P_{y}\left(\frac{1}{2}\right)-P_{y}\left(-\frac{1}{2}\right)\right]\right\} \nonumber \\
 & +\frac{\sqrt{2}}{24}\left\{ 27\left[P_{y+z}\left(\frac{3}{2}\right)-P_{y+z}\left(-\frac{3}{2}\right)\right]+\left[P_{y+z}\left(\frac{1}{2}\right)-P_{y+z}\left(-\frac{1}{2}\right)\right]\right\} \nonumber \\
 & +\frac{\sqrt{2}}{24}\left\{ 27\left[P_{y-z}\left(\frac{3}{2}\right)-P_{y-z}\left(-\frac{3}{2}\right)\right]+\left[P_{y-z}\left(\frac{1}{2}\right)-P_{y-z}\left(-\frac{1}{2}\right)\right]\right\} ,\\
S_{LTT}^{xx}= & \frac{\sqrt{2}}{12}\left\{ 27\left[P_{x+z}\left(\frac{3}{2}\right)-P_{x+z}\left(-\frac{3}{2}\right)\right]+\left[P_{x+z}\left(\frac{1}{2}\right)-P_{x+z}\left(-\frac{1}{2}\right)\right]\right\} \nonumber \\
 & -\frac{\sqrt{2}}{12}\left\{ 27\left[P_{x-z}\left(\frac{3}{2}\right)-P_{x-z}\left(-\frac{3}{2}\right)\right]+\left[P_{x-z}\left(\frac{1}{2}\right)-P_{x-z}\left(-\frac{1}{2}\right)\right]\right\} \nonumber \\
 & -\frac{\sqrt{2}}{12}\left\{ 27\left[P_{y+z}\left(\frac{3}{2}\right)-P_{y+z}\left(-\frac{3}{2}\right)\right]+\left[P_{y+z}\left(\frac{1}{2}\right)-P_{y+z}\left(-\frac{1}{2}\right)\right]\right\} \nonumber \\
 & +\frac{\sqrt{2}}{12}\left\{ 27\left[P_{y-z}\left(\frac{3}{2}\right)-P_{y-z}\left(-\frac{3}{2}\right)\right]+\left[P_{y-z}\left(\frac{1}{2}\right)-P_{y-z}\left(-\frac{1}{2}\right)\right]\right\} ,\\
S_{LTT}^{xy}= & \frac{\sqrt{3}}{16}\left\{ 27\left[P_{x+y+z}\left(\frac{3}{2}\right)-P_{x+y+z}\left(-\frac{3}{2}\right)\right]+\left[P_{x+y+z}\left(\frac{1}{2}\right)-P_{x+y+z}\left(-\frac{1}{2}\right)\right]\right\} \nonumber \\
 & -\frac{\sqrt{3}}{16}\left\{ 27\left[P_{x-y+z}\left(\frac{3}{2}\right)-P_{x-y+z}\left(-\frac{3}{2}\right)\right]+\left[P_{x-y+z}\left(\frac{1}{2}\right)-P_{x-y+z}\left(-\frac{1}{2}\right)\right]\right\} \nonumber \\
 & -\frac{\sqrt{3}}{16}\left\{ 27\left[P_{x+y-z}\left(\frac{3}{2}\right)-P_{x+y-z}\left(-\frac{3}{2}\right)\right]+\left[P_{x+y-z}\left(\frac{1}{2}\right)-P_{x+y-z}\left(-\frac{1}{2}\right)\right]\right\} \nonumber \\
 & +\frac{\sqrt{3}}{16}\left\{ 27\left[P_{x-y-z}\left(\frac{3}{2}\right)-P_{x-y-z}\left(-\frac{3}{2}\right)\right]+\left[P_{x-y-z}\left(\frac{1}{2}\right)-P_{x-y-z}\left(-\frac{1}{2}\right)\right]\right\} \\
S_{TTT}^{xxx}= & \frac{1}{4}\left\{ 27\left[P_{x}\left(\frac{3}{2}\right)-P_{x}\left(-\frac{3}{2}\right)\right]+\left[P_{x}\left(\frac{1}{2}\right)-P_{x}\left(-\frac{1}{2}\right)\right]\right\} \nonumber \\
 & -\frac{\sqrt{2}}{8}\left\{ 27\left[P_{x+y}\left(\frac{3}{2}\right)-P_{x+y}\left(-\frac{3}{2}\right)\right]+\left[P_{x+y}\left(\frac{1}{2}\right)-P_{x+y}\left(-\frac{1}{2}\right)\right]\right\} \nonumber \\
 & -\frac{\sqrt{2}}{8}\left\{ 27\left[P_{x-y}\left(\frac{3}{2}\right)-P_{x-y}\left(-\frac{3}{2}\right)\right]+\left[P_{x-y}\left(\frac{1}{2}\right)-P_{x-y}\left(-\frac{1}{2}\right)\right]\right\} ,\\
S_{TTT}^{yxx}= & -\frac{1}{4}\left\{ 27\left[P_{y}\left(\frac{3}{2}\right)-P_{y}\left(-\frac{3}{2}\right)\right]+\left[P_{y}\left(\frac{1}{2}\right)-P_{y}\left(-\frac{1}{2}\right)\right]\right\} \nonumber \\
 & +\frac{\sqrt{2}}{8}\left\{ 27\left[P_{y+x}\left(\frac{3}{2}\right)-P_{y+x}\left(-\frac{3}{2}\right)\right]+\left[P_{y+x}\left(\frac{1}{2}\right)-P_{y+x}\left(-\frac{1}{2}\right)\right]\right\} \nonumber \\
 & +\frac{\sqrt{2}}{8}\left\{ 27\left[P_{y-x}\left(\frac{3}{2}\right)-P_{y-x}\left(-\frac{3}{2}\right)\right]+\left[P_{y-x}\left(\frac{1}{2}\right)-P_{y-x}\left(-\frac{1}{2}\right)\right]\right\} .\label{eq:STTyxx_interpretation}
\end{align}

The domains for these polarization are given by,
\begin{align}
 & S_{L},S_{T}^{x},S_{T}^{y}\in[-\frac{3}{2},\frac{3}{2}],\nonumber\\
 & S_{LL}\in[-1,1],\quad S_{LT}^{x},S_{LT}^{y},S_{TT}^{xy},S_{TT}^{xx}\in[-\sqrt{3},\sqrt{3}],\nonumber\\
 & S_{LLL}\in[-\frac{9}{10},\frac{9}{10}],\quad S_{LLT}^{x},S_{LLT}^{y}\in[-\frac{3+\sqrt{21}}{10},\frac{3+\sqrt{21}}{10}],\nonumber\\
 & S_{LTT}^{xx},S_{LTT}^{xy}\in[-\sqrt{3},\sqrt{3}],\quad S_{TTT}^{xxx},S_{TTT}^{yxx}\in[-3,3].
\label{eq:spin_range}
\end{align}

The total degree of polarization  is given by,
\begin{align}
d&=\frac{1}{\sqrt{2 s}} \sqrt{(2 s+1) {\rm Tr}[\rho^{2}]-1},\nonumber\\
&=\frac{1}{3}\bigg\{ \frac{12}{5} \Big[(S_L)^2+(S_T^x)^2+(S_T^y)^2 \Big] 
+ \Big[3(S_{LL})^2+(S_{LT}^x)^2+(S_{LT}^y)^2+(S_{TT}^{xx})^2+(S_{TT}^{xy})^2\Big] \nonumber\\ 
&\quad +\frac{1}{3}\Big[20(S_{LLL})^2+30\left((S_{LLT}^x)^2+(S_{LLT}^y)^2\right)  + 3\left((S_{LTT}^{xx})^2+(S_{LTT}^{xy})^2\right)+2\left((S_{TTT}^{xxx})^2+(S_{TTT}^{yxx})^2\right)\Big]\bigg\}^\frac{1}{2}.
\end{align}
Its value ranges between 0 and 1.

\section{Spin-3/2 basis matrices}
\label{C.Matrices}

In order to establish a one-to-one correspondence between the spin components $S_{0},S_{1},...,S_{15}$ in the helicity formalism decomposition and the spin components $1,S_{L},S_{x},...,S_{TTT}^{yxx}$ in the spin density matrix representation, as shown in Table~\ref{tab:correspondence}, we follow the spin basis matrix selection method detailed in Ref.~\cite{Zhang:2023wmd}. Based on Eqs.~\eqref{eq:spin_vector}-\eqref{eq:spin_tensor2},we present the matrix representation of $\Sigma_{\mu}$ as follows,
\begin{align}
\Sigma_{0}=&\frac{1}{4}\bm{1},\quad\Sigma_{1}=\frac{1}{5}\Sigma^{z},\quad\Sigma_{2}=\frac{1}{5}\Sigma^{x},\quad\Sigma_{3}=\frac{1}{5}\Sigma^{y},\nonumber\\
\Sigma_{4}=&\frac{1}{4}\Sigma^{zz},\quad\Sigma_{5}=\frac{1}{6}\Sigma^{xz},\quad\Sigma_{6}=\frac{1}{6}\Sigma^{yz}\nonumber\\
\Sigma_{7}=&\frac{1}{12}\left(\Sigma^{xx}-\Sigma^{yy}\right),\quad\Sigma_{8}=\frac{1}{6}\Sigma^{xy},\quad\Sigma_{9}=\frac{5}{9}\Sigma^{zzz},\nonumber\\
\Sigma_{10}=&\frac{5}{6}\Sigma^{xzz},\quad\Sigma_{11}=\frac{5}{6}\Sigma^{yzz},\quad\Sigma_{12}=\frac{1}{6}\left(\Sigma^{xxz}-\Sigma^{yyz}\right),\nonumber\\
\Sigma_{13}=&\frac{1}{3}\Sigma^{xyz},\Sigma_{14}=\frac{1}{18}\left(\Sigma^{xxx}-3\Sigma^{xyy}\right),\quad\Sigma_{15}=\frac{1}{18}\left(3\Sigma^{xxy}-\Sigma^{yyy}\right).
\label{eq:baisis_matrix}
\end{align}

For convenience, we also provide the explicit expressions for the matrices,
\begin{align}
\Sigma_{0}= & \frac{1}{4}\left(\begin{array}{cccc}
1 & 0 & 0 & 0\\
0 & 1 & 0 & 0\\
0 & 0 & 1 & 0\\
0 & 0 & 0 & 1
\end{array}\right),\Sigma_{1}=\frac{1}{10}\left(\begin{array}{cccc}
3 & 0 & 0 & 0\\
0 & 1 & 0 & 0\\
0 & 0 & -1 & 0\\
0 & 0 & 0 & -3
\end{array}\right),\quad\Sigma_{2}=\frac{1}{10}\left(\begin{array}{cccc}
0 & \sqrt{3} & 0 & 0\\
\sqrt{3} & 0 & 2 & 0\\
0 & 2 & 0 & \sqrt{3}\\
0 & 0 & \sqrt{3} & 0
\end{array}\right),\nonumber\\
\Sigma_{3}= & \frac{i}{10}\left(\begin{array}{cccc}
0 & -\sqrt{3} & 0 & 0\\
\sqrt{3} & 0 & -2 & 0\\
0 & 2 & 0 & -\sqrt{3}\\
0 & 0 & \sqrt{3} & 0
\end{array}\right),\quad\Sigma_{4}=\frac{1}{4}\left(\begin{array}{cccc}
1 & 0 & 0 & 0\\
0 & -1 & 0 & 0\\
0 & 0 & -1 & 0\\
0 & 0 & 0 & 1
\end{array}\right),\quad\Sigma_{5}=\frac{\sqrt{3}}{12}\left(\begin{array}{cccc}
0 & 1 & 0 & 0\\
1 & 0 & 0 & 0\\
0 & 0 & 0 & -1\\
0 & 0 & -1 & 0
\end{array}\right),\nonumber\\
\Sigma_{6}= & \frac{i\sqrt{3}}{12}\left(\begin{array}{cccc}
0 & -1 & 0 & 0\\
1 & 0 & 0 & 0\\
0 & 0 & 0 & 1\\
0 & 0 & -1 & 0
\end{array}\right),\quad\Sigma_{7}=\frac{\sqrt{3}}{12}\left(\begin{array}{cccc}
0 & 0 & 1 & 0\\
0 & 0 & 0 & 1\\
1 & 0 & 0 & 0\\
0 & 1 & 0 & 0
\end{array}\right),\quad\Sigma_{8}=\frac{i\sqrt{3}}{12}\left(\begin{array}{cccc}
0 & 0 & -1 & 0\\
0 & 0 & 0 & -1\\
1 & 0 & 0 & 0\\
0 & 1 & 0 & 0
\end{array}\right),\nonumber\\
\Sigma_{9}= & \frac{1}{6}\left(\begin{array}{cccc}
1 & 0 & 0 & 0\\
0 & -3 & 0 & 0\\
0 & 0 & 3 & 0\\
0 & 0 & 0 & -1
\end{array}\right),\quad\Sigma_{10}=\frac{\sqrt{3}}{6}\left(\begin{array}{cccc}
0 & 1 & 0 & 0\\
1 & 0 & -\sqrt{3} & 0\\
0 & -\sqrt{3} & 0 & 1\\
0 & 0 & 1 & 0
\end{array}\right),\quad\Sigma_{11}=\frac{i\sqrt{3}}{6}\left(\begin{array}{cccc}
0 & -1 & 0 & 0\\
1 & 0 & \sqrt{3} & 0\\
0 & -\sqrt{3} & 0 & -1\\
0 & 0 & 1 & 0
\end{array}\right),\nonumber\\
\Sigma_{12}= & \frac{\sqrt{3}}{12}\left(\begin{array}{cccc}
0 & 0 & 1 & 0\\
0 & 0 & 0 & -1\\
1 & 0 & 0 & 0\\
0 & -1 & 0 & 0
\end{array}\right),\quad\Sigma_{13}=\frac{i\sqrt{3}}{12}\left(\begin{array}{cccc}
0 & 0 & -1 & 0\\
0 & 0 & 0 & 1\\
1 & 0 & 0 & 0\\
0 & -1 & 0 & 0
\end{array}\right),\quad\Sigma_{14}=\frac{1}{6}\left(\begin{array}{cccc}
0 & 0 & 0 & 1\\
0 & 0 & 0 & 0\\
0 & 0 & 0 & 0\\
1 & 0 & 0 & 0
\end{array}\right),\nonumber\\
\Sigma_{15}= & \frac{i}{6}\left(\begin{array}{cccc}
0 & 0 & 0 & -1\\
0 & 0 & 0 & 0\\
0 & 0 & 0 & 0\\
1 & 0 & 0 & 0
\end{array}\right).
\label{eq:basis_matrix2}
\end{align}

\section{$\Omega^{-}\bar{\Omega}^{+}$ polarization correlations matrix}\label{D.spin_correlations}

In Sec.~\ref{s.single} and Sec.~\ref{s.double}, we present the polarization correlation matrix $S_{\mu\nu}$ using our parametrization scheme. For the convenience of adopting alternative parametrization schemes, we provide the expressions of the polarization correlation matrix in terms of  the helicity amplitudes $H_{1}$, $H_{2}$, $H_{3}$, and $H_{4}$.

For the single-tag $\Omega^{-}$, the polarization coefficients are given by~\cite{Zhang:2023wmd},
\begin{align}
S_{0} =& 2\sin^{2}\theta_{\Omega^{-}}\left(\left|H_{1}\right|^{2}+\left|H_{4}\right|^{2}\right)+\left(1+\cos^{2}\theta_{\Omega^{-}}\right)\left(\left|H_{2}\right|^{2}+2\left|H_{3}\right|^{2}\right),\label{eq:S0I}\\
S_{3} =& \frac{1}{\sqrt{2}}\sin2\theta_{\Omega^{-}}\left(2\text{Im}\left[H_{2}H_{1}^{*}\right]+\sqrt{3}\text{Im}\left[H_{3}\left(H_{1}^{*}+H_{4}^{*}\right)\right]\right),\\
S_{4} =& 2\sin^{2}\theta_{\Omega^{-}}\left(\left|H_{4}\right|^{2}-\left|H_{1}\right|^{2}\right)-\left(1+\cos^{2}\theta_{\Omega^{-}}\right)\left|H_{2}\right|^{2},\\
S_{5} =& \sqrt{6}\sin2\theta_{\Omega^{-}}\text{Re}\left[\left(H_{4}-H_{1}\right)H_{3}^{*}\right],\\
S_{7} =& 2\sqrt{3}\sin^{2}\theta_{\Omega^{-}}\text{Re}\left[H_{2}H_{3}^{*}\right],\\
S_{11} =& \frac{\sqrt{2}}{5}\sin2\theta_{\Omega}\left(3\text{Im}\left[H_{1}H_{2}^{*}\right]+\sqrt{3}\text{Im}\left[H_{3}\left(H_{1}^{*}+H_{4}^{*}\right)\right]\right),\\
S_{13} =& 2\sqrt{3}\sin^{2}\theta_{\Omega}\text{Im}\left[H_{2}H_{3}^{*}\right].\label{eq:S13I}
\end{align}

For the double-tag $\Omega^{-}\bar{\Omega}^{+}$, we only present the independent terms, while the dependent terms can be obtained from Eqs.~\eqref{eq:independent1},~\eqref{eq:independent2},~\eqref{eq:independent3} -\eqref{eq:independent4},~\eqref{eq:independent5}-\eqref{eq:independent6},~\eqref{eq:independent7}-\eqref{eq:independent8} as discussed in Sec.~\ref{s.double}.

For the independent terms in the diagonal elements, as show in Eqs.~\eqref{Smunu1}-\eqref{Smunu2}, they are given by,
\begin{align}
S_{0,0} = & 2\sin^{2}\theta\left(\left|H_{1}\right|^{2}+\left|H_{4}\right|^{2}\right)+\left(1+\cos^{2}\theta\right)\left(\left|H_{2}\right|^{2}+2\left|H_{3}\right|^{2}\right),\\
S_{1,1} = & \frac{1}{2}\sin^{2}\theta\left(\left|H_{1}\right|^{2}+9\left|H_{4}\right|^{2}\right)-\frac{1}{4}\left(1+\cos^{2}\theta\right)\left(\left|H_{2}\right|^{2}-6\left|H_{3}\right|^{2}\right),\\
S_{2,2} = &\frac{1}{2}\sin^{2}\theta\left(4\left|H_{1}\right|^{2}+2\left|H_{2}\right|^{2}+3\left|H_{3}\right|^{2}+6\text{Re}\left[H_{1}H_{4}^{*}\right]\right)\nonumber\\
          &+2\sqrt{3}\left(1+\cos^{2}\theta\right)\text{Re}\left[H_{2}H_{3}^{*}\right],\\
S_{3,3} = &-\frac{1}{2}\sin^{2}\theta\left(4\left|H_{1}\right|^{2}-2\left|H_{2}\right|^{2}-3\left|H_{3}\right|^{2}+6\text{Re}\left[H_{1}H_{4}^{*}\right]\right)\nonumber\\
          &-2\sqrt{3}\left(1+\cos^{2}\theta\right)\text{Re}\left[H_{2}H_{3}^{*}\right],\\
S_{4,4} = & 2\sin^{2}\theta\left(\left|H_{1}\right|^{2}+\left|H_{4}\right|^{2}\right)+\left(1+\cos^{2}\theta\right)\left(\left|H_{2}\right|^{2}-2\left|H_{3}\right|^{2}\right),\\
S_{5,5} = & 6\sin^{2}\theta\left(\left|H_{3}\right|^{2}+2\text{Re}\left[H_{1}H_{4}^{*}\right]\right),\\
S_{6,6} = & 6\sin^{2}\theta\left(\left|H_{3}\right|^{2}-2\text{Re}\left[H_{1}H_{4}^{*}\right]\right),\\
S_{7,7} = & 12\sin^{2}\theta\text{Re}\left[H_{1}H_{4}^{*}\right]+6\left(1+\cos^{2}\theta\right)\left|H_{3}\right|^{2},\\
S_{9,9} = & \frac{9}{50}\sin^{2}\theta\left(9\left|H_{1}\right|^{2}+\left|H_{4}\right|^{2}\right)-\frac{27}{100}\left(1+\cos^{2}\theta\right)\left(3\left|H_{2}\right|^{2}+2\left|H_{3}\right|^{2}\right),\\
S_{10,10} = & \frac{3}{25}\sin^{2}\theta\left(6\left|H_{1}\right|^{2}+3\left|H_{2}\right|^{2}+2\left|H_{3}\right|^{2}+4\text{Re}\left[H_{1}H_{4}^{*}\right]\right)\nonumber\\
            &-\frac{12\sqrt{3}}{25}\left(1+\cos^{2}\theta\right)\text{Re}\left[H_{2}H_{3}^{*}\right],\\
S_{11,11} = & -\frac{3}{25}\sin^{2}\theta\left(6\left|H_{1}\right|^{2}-3\left|H_{2}\right|^{2}-2\left|H_{3}\right|^{2}+4\text{Re}\left[H_{1}H_{4}^{*}\right]\right)\nonumber\\
            &+\frac{12\sqrt{3}}{25}\left(1+\cos^{2}\theta\right)\text{Re}\left[H_{2}H_{3}^{*}\right],\\
S_{12,12} = & 12\sin^{2}\theta\text{Re}\left[H_{1}H_{4}^{*}\right]-6\left(1+\cos^{2}\theta\right)\left|H_{3}\right|^{2},\\
S_{14,14} = & 18\sin^{2}\theta\left|H_{4}\right|^{2}.
\end{align}

For the independent terms shown in Eqs.~\eqref{Smunu0_1}-\eqref{Smunu0_2}, they are given by,
\begin{align}
S_{4,0} = & -2\sin^{2}\theta\left(\left|H_{1}\right|^{2}-\left|H_{4}\right|^{2}\right)-\left(1+\cos^{2}\theta\right)\left|H_{2}\right|^{2},\\
S_{9,1} = & -\frac{9}{10}\sin^{2}\theta\left(\left|H_{1}\right|^{2}-\left|H_{4}\right|^{2}\right)+\frac{3}{20}\left(1+\cos^{2}\theta\right)\left(3\left|H_{2}\right|^{2}-8\left|H_{3}\right|^{2}\right),\\
S_{14,2} = & 3\left|H_{3}\right|^{2}\sin^{2}\theta.
\end{align}

For the independent terms shown in Eqs.~\eqref{Smunu1_1}-\eqref{Smunu1_2}, they are given by,
\begin{align}
S_{7,0} = & 2\sqrt{3}\text{Re}\left[H_{2}H_{3}^{*}\right]\sin^{2}\theta,\\
S_{13,0} = & 2\sqrt{3}\text{Im}\left[H_{2}H_{3}^{*}\right]\sin^{2}\theta,\\
S_{13,7} = & 12\text{Im}\left[H_{1}H_{4}^{*}\right]\sin^{2}\theta,\\
S_{7,5} = & 3\sqrt{2}\text{Re}\left[H_{2}H_{4}^{*}\right]\sin2\theta,\\
S_{13,5} = & 3\sqrt{2}\text{Im}\left[H_{2}H_{4}^{*}\right]\sin2\theta,\\
S_{15,7} = & 3\sqrt{6}\text{Im}\left[H_{3}H_{4}^{*}\right]\sin2\theta,\\
S_{14,12} = & 3\sqrt{6}\text{Re}\left[H_{3}H_{4}^{*}\right]\sin2\theta.
\end{align}

For the independent terms shown in Eqs.~\eqref{Smunu2_1}-\eqref{Smunu2_2}, they are given by,
\begin{align}
S_{5,0}= & -\sqrt{6}\text{Re}\left[H_{3}\left(H_{1}^{*}-H_{4}^{*}\right)\right]\sin2\theta,\\
S_{6,1} = & \frac{\sqrt{6}}{2}\text{Im}\left[H_{3}\left(H_{1}^{*}+3H_{4}^{*}\right)\right]\sin2\theta,\\
S_{5,4} = & \sqrt{6}\text{Re}\left[H_{3}\left(H_{1}^{*}+H_{4}^{*}\right)\right]\sin2\theta,\\
S_{9,6} = & \frac{3\sqrt{6}}{10}\text{Im}\left[H_{3}\left(3H_{1}^{*}-H_{4}^{*}\right)\right]\sin2\theta,\\
S_{8,2} = & \frac{\sqrt{2}}{2}\sin2\theta\left(3\text{Im}\left[H_{2}H_{4}^{*}\right]-2\sqrt{3}\text{Im}\left[H_{1}H_{3}^{*}\right]\right),\\
S_{12,2} = & \frac{\sqrt{2}}{2}\sin2\theta\left(3\text{Re}\left[H_{2}H_{4}^{*}\right]-2\sqrt{3}\text{Re}\left[H_{1}H_{3}^{*}\right]\right),\\
S_{11,7} = & -\frac{3\sqrt{2}}{5}\sin2\theta\left(\sqrt{3}\text{Im}\left[H_{1}H_{3}^{*}\right]+\text{Im}\left[H_{2}H_{4}^{*}\right]\right),\\
S_{12,10} = & \frac{3}{5}\sqrt{2}\sin2\theta\left(\sqrt{3}\text{Re}\left[H_{1}H_{3}^{*}\right]+\text{Re}\left[H_{2}H_{4}^{*}\right]\right),\\
S_{6,2} = & 6\text{Im}\left[H_{1}H_{4}^{*}\right]\sin^{2}\theta+2\sqrt{3}\left(1+\cos^{2}\theta\right)\text{Im}\left[H_{2}H_{3}^{*}\right],\\
S_{11,5} = & \frac{12}{5}\text{Im}\left[H_{1}H_{4}^{*}\right]\sin^{2}\theta-\frac{6\sqrt{3}}{5}\left(1+\cos^{2}\theta\right)\text{Im}\left[H_{2}H_{3}^{*}\right],\\
S_{10,2} = & -\frac{3}{5}\sin^{2}\theta\left(2\left|H_{1}\right|^{2}+\left|H_{2}\right|^{2}-\left|H_{3}\right|^{2}-2\text{Re}\left[H_{1}H_{4}^{*}\right]\right)\nonumber\\
           &-\frac{\sqrt{3}}{5}\left(1+\cos^{2}\theta\right)\text{Re}\left[H_{2}H_{3}^{*}\right],\\
S_{11,3} = & \frac{3}{5}\sin^{2}\theta\left(2\left|H_{1}\right|^{2}-\left|H_{2}\right|^{2}+\left|H_{3}\right|^{2}-2\text{Re}\left[H_{1}H_{4}^{*}\right]\right)\nonumber\\
           &+\frac{\sqrt{3}}{5}\text{Re}\left[H_{2}H_{3}^{*}\right]\left(1+\cos^{2}\theta\right),
\end{align}

For the independent terms shown in Eqs.~\eqref{Smunu3_1}-\eqref{Smunu3_2}, they are given by,
\begin{align}
S_{3,0} = & -\frac{\sqrt{2}}{2}\sin2\theta\left(2\text{Im}\left[H_{1}H_{2}^{*}\right]-\sqrt{3}\text{Im}\left[H_{3}\left(H_{1}^{*}+H_{4}^{*}\right)\right]\right),\\
S_{11,0} = & \frac{\sqrt{2}}{5}\sin2\theta\left(3\text{Im}\left[H_{1}H_{2}^{*}\right]+\sqrt{3}\text{Im}\left[H_{3}\left(H_{1}^{*}+H_{4}^{*}\right)\right]\right),\\
S_{2,1} = & \frac{\sqrt{2}}{4}\sin2\theta\left(2\text{Re}\left[H_{1}H_{2}^{*}\right]-\sqrt{3}\text{Re}\left[H_{3}\left(H_{1}^{*}-3H_{4}^{*}\right)\right]\right),\\
S_{10,1} = & -\frac{\sqrt{2}}{10}\sin2\theta\left(3\text{Re}\left[H_{1}H_{2}^{*}\right]+\sqrt{3}\text{Re}\left[H_{3}\left(H_{1}^{*}-3H_{4}^{*}\right)\right]\right),\\
S_{9,2} = & \frac{3\sqrt{2}}{20}\sin2\theta\left(6\text{Re}\left[H_{1}H_{2}^{*}\right]-\sqrt{3}\text{Re}\left[H_{3}\left(3H_{1}^{*}+H_{4}^{*}\right)\right]\right),\\
S_{4,3} = & -\frac{\sqrt{2}}{2}\sin2\theta\left(2\text{Im}\left[H_{1}H_{2}^{*}\right]-\sqrt{3}\text{Im}\left[H_{3}\left(H_{1}^{*}-H_{4}^{*}\right)\right]\right),\\
S_{11,4} = & -\frac{\sqrt{2}}{5}\sin2\theta\left(3\text{Im}\left[H_{1}H_{2}^{*}\right]+\sqrt{3}\text{Im}\left[H_{3}\left(H_{1}^{*}-H_{4}^{*}\right)\right]\right),\\
S_{10,9} = & \frac{3\sqrt{2}}{50}\sin2\theta\left(9\text{Re}\left[H_{1}H_{2}^{*}\right]+\sqrt{3}\text{Re}\left[H_{3}\left(3H_{1}^{*}+H_{4}^{*}\right)\right]\right).
\end{align}

\section{spin transfer matrices}\label{E.spin transfer}

Based on the parametrization schemes introduced in Sec.~\ref{s.decay}, we derive the specific expressions for the polarization transfer matrix $a_{\mu\nu}$ and $b_{\mu\nu}$.  For the coefficients of $a_{\mu\nu}$, 14 out of 16 are non-zero, and these non-zero coefficients are represented as follows~\cite{Perotti:2018wxm},
\begin{align}
a_{0,0} = & 1,\\
a_{0,3} = & \alpha_{D},\\
a_{1,0} = & \alpha_{D}\sin\theta\cos\phi,\\
a_{1,1} = & \gamma_{D}\cos\theta\cos\phi-\beta_{D}\sin\phi,\\
a_{1,2} = & -\beta_{D}\cos\theta\cos\phi-\gamma_{D}\sin\phi,\\
a_{1,3} = & \sin\theta\cos\phi,\\
a_{2,0} = & \alpha_{D}\sin\theta\sin\phi,\\
a_{2,1} = & \beta_{D}\cos\phi+\gamma_{D}\cos\theta\sin\phi,\\
a_{2,2} = & \gamma_{D}\cos\phi-\beta_{D}\cos\theta\sin\phi,\\
a_{2,3} = & \sin\theta\sin\phi,\\
a_{3,0} = & \alpha_{D}\cos\theta,\\
a_{3,1} = & -\gamma_{D}\sin\theta,\\
a_{3,2} = & \beta_{D}\sin\theta,\\
a_{3,3} = & \cos\theta.
\end{align}

For the coefficients $b_{\mu\nu}$, 52 out of 64 are non-zero.  While partial expressions are provided in Ref.~\cite{Perotti:2018wxm}, our unique basis matrix selection requires a different formulation. We systematically present these expressions, identifying 36 independent coefficients as follows,
\begin{align}
b_{0,0} = &1,\\
b_{1,1} = & -\frac{4}{5}\gamma_{D}\sin\theta,\\
b_{1,2} = & \frac{4}{5}\beta_{D}\sin\theta,\\
b_{1,3} = & \frac{2}{5}\cos\theta,\\
b_{2,1} = & -\frac{4}{5}\left(-\gamma_{D}\cos\theta\cos\phi+\beta_{D}\sin\phi\right),\\
b_{2,2} = & -\frac{4}{5}\left(\beta_{D}\cos\theta\cos\phi+\gamma_{D}\sin\phi\right),\\
b_{2,3} = & \frac{2}{5}\sin\theta\cos\phi,\\
b_{3,1} = &\frac{4}{5}\left(\beta_{D}\cos\phi+\gamma_{D}\cos\theta\sin\phi\right),\\
b_{3,2} = &\frac{4}{5}\left(\gamma_{D}\cos\phi-\beta_{D}\cos\theta\sin\phi\right),\\
b_{3,3} = &\frac{2}{5}\sin\theta\sin\phi,\\
b_{4,0} = &-\frac{1}{4}\left(1+3\cos2\theta\right),\\
b_{5,0} = &-\sin\theta\cos\theta\cos\phi,\\
b_{6,0} = & -\sin\theta\cos\theta\sin\phi,\\
b_{7,0} = &-\frac{1}{2}\sin^{2}\theta\cos2\phi,\\
b_{8,0} = & -\sin^{2}\theta\sin\phi\cos\phi,\\
b_{9,1} = & \frac{1}{4}\gamma_{D}\left(\sin\theta+5\sin3\theta\right),\\
b_{9,2} = & -\frac{1}{4}\beta_{D}\left(\sin\theta+5\sin3\theta\right),\\
b_{9,3} = & -\frac{1}{4}\left(3\cos\theta+5\cos3\theta\right),\\
b_{10,1} = & \frac{1}{8}\left[2\beta_{D}\left(3+5\cos2\theta\right)\sin\phi-\gamma_{D}\left(\cos\theta+15\cos3\theta\right)\cos\phi\right],\\
b_{10,2} = & \frac{1}{8}\left[2\gamma_{D}\left(3+5\cos2\theta\right)\sin\phi+\beta_{D}\left(\cos\theta+15\cos3\theta\right)\cos\phi\right],\\
b_{10,3} = & -\frac{3}{8}\left(\sin\theta+5\sin3\theta\right)\cos\phi,\\
b_{11,1} = &-\frac{1}{8}\left[2\beta_{D}\left(3+5\cos2\theta\right)\cos\phi+\gamma_{D}\left(\cos\theta+15\cos3\theta\right)\sin\phi\right],\\
b_{11,2} = &-\frac{1}{8}\left[2\gamma_{D}\left(3+5\cos2\theta\right)\cos\phi-\beta_{D}\left(\cos\theta+15\cos3\theta\right)\sin\phi\right],\\
b_{11,3} = &-\frac{3}{8}\left(\sin\theta+5\sin3\theta\right)\sin\phi,\\
b_{12,1} = & \frac{1}{4}\sin\theta\left[4\beta_{D}\cos\theta\sin2\phi-\gamma_{D}\left(1+3\cos2\theta\right)\cos2\phi\right],\\
b_{12,2} = & \frac{1}{4}\sin\theta\left[4\gamma_{D}\cos\theta\sin2\phi+\beta_{D}\left(1+3\cos2\theta\right)\cos2\phi\right],\\
b_{12,3} = & -\frac{3}{2}\sin^{2}\theta\cos\theta\cos2\phi,\\
b_{13,1} = &-\frac{1}{4}\sin\theta\left[4\beta_{D}\cos\theta\cos2\phi+\gamma_{D}\left(1+3\cos2\theta\right)\sin2\phi\right],\\
b_{13,2} = &-\frac{1}{4}\sin\theta\left[4\gamma_{D}\cos\theta\cos2\phi-\beta_{D}\left(1+3\cos2\theta\right)\sin2\phi\right],\\
b_{13,3} = &-3\sin^{2}\theta\cos\theta\sin\phi\cos\phi,\\
b_{14,1} = & \frac{1}{2}\sin^{2}\theta\left(\beta_{D}\sin3\phi-\gamma_{D}\cos\theta\cos3\phi\right),\\
b_{14,2} = & \frac{1}{2}\sin^{2}\theta\left(\gamma_{D}\sin3\phi+\beta_{D}\cos\theta\cos3\phi\right),\\
b_{14,3} = & -\frac{1}{2}\sin^{3}\theta\cos3\phi,\\
b_{15,1} = & -\frac{1}{2}\sin^{2}\theta\left(\beta_{D}\cos3\phi+\gamma_{D}\cos\theta\sin3\phi\right),\\
b_{15,2} = & -\frac{1}{2}\sin^{2}\theta\left(\gamma_{D}\cos3\phi-\beta_{D}\cos\theta\sin3\phi\right),\\
b_{15,3} = & -\frac{1}{2}\sin^{3}\theta\sin3\phi,
\end{align}
Additionally, we present the 16 derived dependent coefficients as
\begin{align}
\left\{ \begin{array}{c}
b_{0,3},b_{1,0},b_{2,0},b_{3,0},b_{4,3},b_{5,3},b_{6,3},b_{7,3}\\
b_{8,3},b_{9,0},b_{10,0},b_{11,0},b_{12,0},b_{13,0},b_{14,0},b_{15,0}
\end{array}\right\}   =\alpha_{D}\left\{ \begin{array}{c}
b_{0,0},b_{1,3},b_{2,3},b_{3,3},b_{4,0},b_{5,0},b_{6,0},b_{7,0}\\
b_{8,0},b_{9,3},b_{10,3},b_{11,3},b_{12,3},b_{13,3},b_{14,3},b_{15,3}
\end{array}\right\} .
\end{align}

\bibliography{paper_revision_v2}

\end{document}